\documentclass[
aps,
showpacs,
preprintnumbers,
amsmath,
onecolumn,
amssymb,
superscriptaddress,
prx, 
floatfix
]{revtex4-2}

\usepackage{amsfonts}
\usepackage{amsmath}
\usepackage{amssymb}
\usepackage[export]{adjustbox}
\usepackage{tcolorbox}
\usepackage{eurosym}
\usepackage{graphicx}
\usepackage{dcolumn}
\usepackage{bm}
\usepackage[utf8]{inputenc}
\usepackage{xkcdcolors}
\usepackage{tabularx}
\usepackage{epstopdf}
\usepackage{xcolor}
\usepackage{mathrsfs}
\usepackage{mathtools}
\usepackage{float}
\usepackage{verbatim}
\usepackage{ragged2e}
\usepackage{nccmath}
\usepackage{hyperref}

\begin{document}

\title{Commodity-specific triads in the Dutch inter-industry production network}

\author{Marzio Di Vece}
\email{marzio.divece@imtlucca.it}
\affiliation{IMT School for Advanced Studies Lucca, P.zza San Francesco 19, 55100 Lucca (Italy)}
\affiliation{Lorentz Institute for Theoretical Physics, Leiden University, Niels Bohrweg 2, 2333CA Leiden (The Netherlands)}
\affiliation{Scuola Normale Superiore, P.zza dei Cavalieri 7, Pisa (Italy)}
\author{Frank P. Pijpers}
\affiliation{Statistics Netherlands, Henri Faasdreef 312, 2492 JP Den Haag (the Netherlands)}
\affiliation{Korteweg - de Vries Institute for Mathematics, University of Amsterdam, Amsterdam (the Netherlands)}
\author{Diego Garlaschelli}
\affiliation{IMT School for Advanced Studies Lucca, P.zza San Francesco 19, 55100 Lucca (Italy)}
\affiliation{Lorentz Institute for Theoretical Physics, Leiden University, Niels Bohrweg 2, 2333CA Leiden (The Netherlands)}
\affiliation{INdAM-GNAMPA Istituto Nazionale di Alta Matematica (Italy)}

\date{\today}

\begin{abstract}
Triadic motifs are the smallest building blocks of higher-order interactions in complex networks and can be detected as over-occurrences with respect to null models with only pair-wise interactions. 
Recently, the motif structure of production networks has attracted attention in light of its possible role in the propagation of economic shocks. However, its characterization at the level of individual commodities is still poorly understood.
Here we analyze both binary and weighted triadic motifs in the Dutch inter-industry production network disaggregated at the level of 187 commodity groups, which Statistics Netherlands reconstructed from National Accounts registers, surveys and known empirical data. 
We introduce appropriate null models that filter out node heterogeneity and the strong effects of link reciprocity and find that, while the aggregate network that overlays all products is characterized by a multitude of triadic motifs, most single-product layers feature no significant motif, and roughly $85\%$ of the layers feature only two motifs or less. 
This result paves the way for identifying a simple `triadic fingerprint' of each commodity and for reconstructing most product-specific networks from partial information in a pairwise fashion by controlling for their reciprocity structure. We discuss how these results can help statistical bureaus identify fine-grained information in structural analyses of interest for policymakers. 
\end{abstract}

\maketitle

In the last decade, the increasing availability of data at the industry and firm level led to a vast number of studies analyzing the system of customer-supplier trade relationships - the \textit{production network} - among industries~\cite{acemoglu_network_2012,aobdia_inter-industry_2014,atalay_how_2017,bouakez_transmission_2009,brintrup_predicting_2018,pichler_simultaneous_2022} or firms~\cite{bacilieri_firm_2023,atalay_network_2011,bernard_production_2019,buiten_reconstruction_2021,carvalho_supply_nodate,carvalho_production_2019,mungo_reconstructing_2023,cohen_economic_2008,dhyne_belgian_2015,dhyne_trade_2021,diem_quantifying_2022,cardoza_worker_2020,chacha_mapping_2022,demir_financial_2022,peydro_production_2020,spray_industries_2018,kumar_distress_2021,goto_estimating_2017,hooijmaaijers_methodology_nodate,ialongo_reconstructing_2022,inoue_firm-level_2019,inoue_propagation_2020,kashiwagi_propagation_2021,konig_aggregate_2022,kosasih_machine_2022,maluck_motif_2017,mattsson_functional_2021, McNerney_how_2021,mizuno_structure_2014,ohnishi_network_2010,rachkov_potential_2021,taschereau-dumouchel_cascades_2017,WATANABE2013741} and their impact on country-level macroeconomic statistics~\cite{diem_estimating_2023}.

The heterogeneity encoded in the production network structure plays an essential role in amplifying economic growth~\cite{McNerney_how_2021} and in the propagation of shocks~\cite{acemoglu_network_2012,carvalho_production_2019} related to exogenous events, such as Hurricane Sandy~\cite{kashiwagi_propagation_2021}, the Great East Asian Earthquake~\cite{carvalho_supply_nodate, inoue_firm-level_2019}, the Covid-19 pandemic~\cite{diem_quantifying_2022, inoue_propagation_2020, pichler_simultaneous_2022}, or endogenous events such as the 2008 financial crisis~\cite{maluck_network_2015,wang_motif_2022}.

Even in the time of globalization - characterized by highly interconnected global supply chains - domestic production networks are still relevant. In fact, 
it has been shown that for a small country as Belgium, while almost all firms directly or indirectly import and export to foreign firms, these exchanges represent the minority of domestic firms' total revenues~\cite{dhyne_trade_2021}. 

While aggregated information about single firms is contained in most National Statistical Institutes' repositories, reliable data on input/output relationships is available only for a small number of countries.
For instance, the Compustat dataset contains the major customers
of the publicly listed firms in the USA~\cite{atalay_network_2011}.
The FactSet Revere dataset contains major customers of publicly listed firms at a global level, with a focus on the USA, Europe, and Asia~\cite{konig_aggregate_2022}.
Two datasets are commercially available in Japan, namely, the dataset collected by Tokyo Shoko Research Ltd. (TSR)\cite{carvalho_supply_nodate} and the one collected by Teikoku DataBank Inc. (TDB)\cite{mizuno_structure_2014}. They are characterized by a high coverage of Japanese firms but with a limited amount of commercial partners.
Other domestic datasets contain transaction values among VAT-liable firms: this is the case for countries such as Brazil~\cite{mungo_reconstructing_2023}, Belgium~\cite{dhyne_belgian_2015}, Hungary~\cite{diem_quantifying_2022}, Ecuador~\cite{bacilieri_firm_2023}, Kenya~\cite{chacha_mapping_2022}, Turkey~\cite{demir_financial_2022}, Spain~\cite{peydro_production_2020}, Rwanda and Uganda~\cite{spray_industries_2018}, West Bengal~\cite{kumar_distress_2021}; or contain transaction values among the totality of registered domestic firms such as in the case of Dominican Republic~\cite{cardoza_worker_2020} and Costa Rica~\cite{alfaro2018costa}.

\begin{figure*}[ht!]
    \centering
\includegraphics[width=1\linewidth]{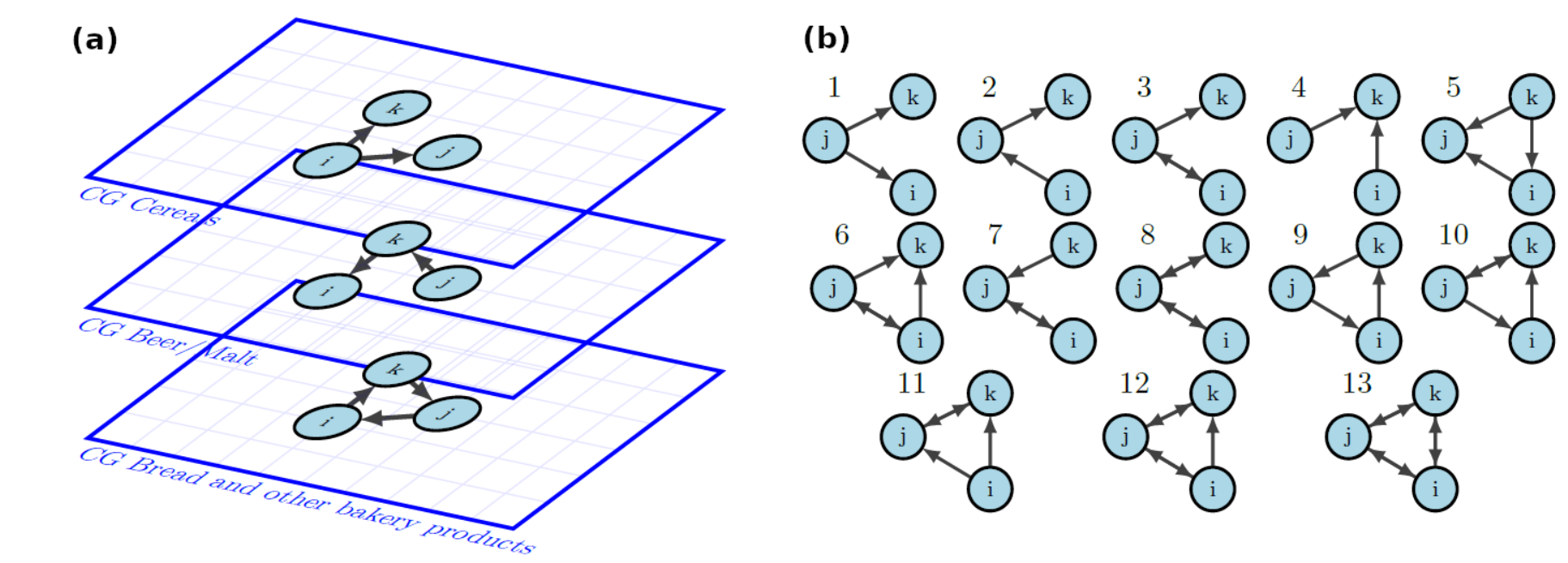}
    \caption{(a) Graphical representation of the Dutch multi-layer production network. For illustrative purposes, we represent three industries/firms $i$, $j$ and $k$ as nodes, all placed on three commodity group layers, namely (from top to bottom): Cereals, Beer/Malt, Bread and other bakery products.
    The connections between the same three nodes are different in the different layers.
    (b) The possible 13 types of connected triadic subgraphs. Each triple of Industries/Firms can trade different products by forming, on each commodity-specific layer, either one of the 13 possible connected subgraphs or one of the remaining subgraphs where at least one node is disconnected (not shown).}
    \label{fig1}
\end{figure*}

However, in production networks, user firms connect to supplier firms to buy goods for their own production. Customer-supplier relationships are, hence, characterized by an intrinsic product granularity that is usually neglected. The importance of product-specific information has been highlighted, for instance, in a rare study that utilizes surveys with limited data for Japanese automotive firms~\cite{kito_how_2015}.
Generally, in the economic theory of industries and firms, the problem of product granularity is `solved' artificially, by assuming that industries/firms supply a single product~\cite{acemoglu_network_2012, bouakez_transmission_2009}.
This is an oversimplification that often conflicts with reality: indeed, a single firm can possess more than a production pipeline and is capable of supplying multiple products (e.g. Samsung, a Telecommunication company, sells also household appliances, and multinational companies such as Amazon and Google supply a large number of different products). 

Recently, Statistics Netherlands (CBS) reconstructed from national statistics two multi-layer production network datasets for domestic intermediate trade of Dutch firms for 2012~\cite{hooijmaaijers_methodology_nodate} and 2018~\cite{buiten_reconstruction_2021}, with each layer corresponding to a different product exchanged by a firm for its own production process, as illustratively depicted in Fig.~\ref{fig1}(a). 
The 2012 dataset has been recently used to prove the complementarity structure of production networks~\cite{mattsson_functional_2021} by inspecting the number of cycles of order 3 and 4 compared to a null model taking into consideration the in-degree and out-degree distributions. 
We use the improved version for 2018 and construct an inter-industry network that will be presented in the next section.

In this study, we focus on triadic motifs and anti-motifs that are over-occurrences and under-occurrences of different patterns of directed triadic connections, respectively. They are represented in Fig.~\ref{fig1}(b). 
Triadic and tetradic connections are known as the building blocks of complex networks~\cite{milo_network_2002}, playing the role of functional modules or evolutionary signs in biological networks~\cite{shen-orr_network_2002,stivala_testing_2021},
homophily-driven connections in social networks~\cite{asikainen_cumulative_2020},
complementarity-driven structures in production networks~\cite{ohnishi_network_2010,mattsson_functional_2021}, their change in time being interpreted as self-organizing processes in the World Trade Web (WTW)~\cite{hutchison_triadic_2012,maratea_triadic_2016}, and early-warning signals of topological collapse in inter-bank networks~\cite{squartini_early-warning_2013, squartini_stationarity_2014} and stock market networks~\cite{wang_motif_2022}.
It has been proven that for the majority of (available) real-world networks, the triadic structure is maximally random~\cite{colomer-de-simon_deciphering_2013} and by fixing it their global structure is statistically determined~\cite{jamakovic_how_2009}.

In contrast, research on weighted motifs and anti-motifs is still underdeveloped. To our knowledge, only one study involves trade volumes circulating on triadic subgraphs, using a probabilistic model based on random walks on the WTW~\cite{picciolo_weighted_2022}.

Motif detection strictly depends not only on the properties of the real network but also on the randomization method used for the computation of random expectations.
In Network Science literature, various methods have been advanced for network randomization, primarily edge-stub methods, edge-swapping methods, and Maximum-Entropy methods, we focus on the latter.
Randomization methods based on Entropy Maximization~\cite{jaynes_information_1,jaynes_information_1957,jaynes_rationale_1982} build Graph Probability distributions that are maximally random by construction. Available global or node-specific data are encoded as constraints in the optimization procedure, and their corresponding Lagrange Multipliers are computed by Maximum Likelihood Estimation (MLE)~\cite{garlaschelli_maximum_2008}.
This theoretical framework has been proven to successfully reconstruct economic and financial systems~\cite{bardoscia_physics_2021,cimini_statistical_2019,cimini_mastrandrea_squartini_2021,squartini_garlaschelli_2017}, statistically predicting both the topology and the weights of the WTW ~\cite{garlaschelli_fitness_2004,squartini_randomizing_2011,squartini_randomizing_2011-1}, in an integrated~\cite{mastrandrea_enhanced_2014}, or conditional fashion~\cite{parisi_faster_2020}, with only structural constraints, or informing the models with economic factors~\cite{Almog_2019, di_vece_gravity_2022, di_vece_reconciling_2023}, statistically predicting banks' risk exposures~\cite{cimini_systemic_2015}, and most recently, statistically reconstructing payment flows among Dutch firms that were clients of ABN Amro Bank or ING Bank, constraining their industry-specific production functions~\cite{ialongo_reconstructing_2022}. 
These methods have been proven to give the best insurance of unbiasedness with respect to missing data, as seen by independent testing~\cite{anand_missing_2018,lebacher_in_2019, RAMADIAH2020103817,mazzarisi_methods_2017}.
Two studies using Maximum-Entropy modeling are especially worthy of note for motif detection: a theoretical study where the authors develop null models for triadic motif detections and compute z-scores of triadic occurrences analytically~\cite{squartini_analytical_2011}, and an applied study where triadic motifs and their time evolution are used as early warnings of topological collapse during the 2008 financial crisis~\cite{squartini_early-warning_2013}.

Our contribution goes in this direction, using Maximum-Entropy methods constraining degree distributions and strength distributions - in their directed form and taking into account their reciprocal nature - to characterize triadic connections and the total money circulating on them for different product layers of the reconstructed Dutch production network. An analysis of this kind can give better insight into how much product-level granularity is needed in production network datasets and how the links and weights of a production network are organized for different products.
Once product layer patterns have been detected, National Bureau officials - having experience in the domestic trade of that single commodity - can infer if such motifs and anti-motifs are due to commodity-specific characteristics, market imbalances, or represent structures aided by laws. If unbalances and anomalies are detected, executive government agencies can use this as input to eventually advance policy laws to nudge a more convenient redistribution of connections and trade volumes.

\section*{Results}

\subsection*{The CBS Production Network}\label{Sec_Data} 

The CBS production network for 2018~\cite{buiten_reconstruction_2021} improves on the 2012 version by integrating more auxiliary micro and industry-level data.
Before going into detail it is helpful to explain which are the industry classifications and product classifications used by Statistics Netherlands. Industries are classified using the Dutch Standard Industrial Classifications, in brief SBI, which are equivalent to the European Standard Classification NACE rev.2 in the first two digits, although the subsequent digits can differ. Statistics Netherlands has industry data on two different levels, the SBI4 level, containing $132$ industries, and the SBI5 level, corresponding to $888$ industries. 
Regarding the CPA product classification, Statistics Netherlands uses a modified version of the original European CPA, mainly at $4$ and $6$ digits. In the data, we retrieved $192$ commodities for 4 digits and $623$ commodities at the 6-digit level.

Firm-level data is obtained from the Statistical Business Register (SBR) for 2018 for over $1\, 700\, 000$ firms. The SBR contains values about net turnover, geographical location, business id, and business sector at the SBI5 classification level.
After cleaning for micro-firms with annual net turnover below $10\, 000$\euro{}, around $900\, 000$ firms remain, accounting for $99.5\%$ of the Dutch economy output in 2018. 
Details regarding the breakdown of output and input at the commodity level are primarily available at the industry level and for a limited number of firms. The Dutch National Supply-Use tables provide data on inter-industry and intra-industry intermediate input/output transactions for various commodities, classified at the Dutch CPA 4-digit level with industries categorized according to the SBI4 level.

While industry-wide transactions are validated, estimating output and input for individual firms per commodity and matching suppliers with users within commodity layers remains a challenge. To estimate supply per firm, domestic turnover, calculated as VAT turnover minus export turnover, is employed as a distributional key. Firms are assumed to supply in proportion to the ratio of their domestic turnover to the overall industry turnover. Additional adjustments are made for wholesale and retail trade firms to account only for domestic turnover associated with actual production.

Estimating use per firm from VAT turnover data involves determining the ratio between intermediate use and turnover. This ratio is estimated using SBS survey data. The breakdown of supply/use per firm at the commodity level is available for a relatively large number of firms through surveys conducted by Structural Business Statistics (SBS) for commercial firms, Prodcom for manufacturing firms, and estimates generated by National Accounts for non-commercial firms.

Specifically, SBS provides a breakdown of sales and intermediate purchases into ten to twenty commodity categories for small firms and at the CPA-classification level for large firms. Prodcom conducts a similar survey. SBS categories are then mapped into CPA commodities by National Account experts.

For firms not covered by the aforementioned surveys, the breakdown in commodities of intermediate supply/use is estimated using the distribution of the industries as a whole from the Supply-Use tables. This approximation can result in implausible values of annual supply and use. To address this issue, thresholds are imposed, setting supply values below $2 \,000$\euro{} and also use values below $1 \,000$\euro{} to zero. Finally, an Iterative Proportional Fitting (IPF) procedure is implemented to ensure consistency with industry-level Tables.

Once supply and use per firm per commodity are obtained, their out-degree distribution is estimated using stylized facts from Japanese firms~\cite{WATANABE2013741}, connecting out-degrees with firm sizes through a power-law function, while their in-degree distribution is estimated assuming a power-law connection to firm-specific input at the commodity-level, an assumption that is consistent with recent studies on Dutch inter-firm payments~\cite{ialongo_reconstructing_2022}.

Once in-degree and out-degree distributions per commodity are estimated for each firm, suppliers and users are matched according to a deterministic procedure that takes into account (1) a company score, encoding their net turnover, (2) a distance score, that takes into account their mutual distance, (3) the presence of a link between respective industries in the Supply-Use tables, (4) the presence of the observed relationship in the Dun \& Bradstreet dataset, i.e. a dataset containing the list of the users of the 500 largest suppliers in the Dutch Economy.
After the computation of the related `link score', users in each commodity layer are ordered according to their purchase volumes. The top user, then, selects the best X suppliers and establish a connection with them, where X represents its commodity-specific in-degree. The procedure continues from the second-highest purchasing volume user to the last until no available links remain and degree distributions are reproduced.
Network weights are then distributed across generated links according to a power-law distribution.
Finally, the resulting weighted inter-firm network at the $650$ commodity level (National CPA level 6) is compared to the Supply-Use tables (National CPA level 4) and consequent adjustments are made to weights and links. Further details can be found in~\cite{buiten_reconstruction_2021}.

We aggregate the inter-firm network at the commodity level, passing from $623$ commodities (CPA level 6) to $192$ commodities (CPA level 4, compatible with Supply-Use tables). Then, we aggregate firms in industries at the SBI5 level, taking their business sector ids from the SBR.
For the topic of interest, the self-loops implied by intra-industry trade are not important and can be removed from the dataset without adversely affecting the subsequent analysis. 
After cleaning for intra-industry trade, we  obtain a multi-layer inter-industry production network containing linkages and weights for $862$ industries (nodes) and $187$ commodity groups (layers).

The firm-level reconstructed dataset is not without limitations. One source of error arises from the breakdown provided by SBS and Prodcom surveys, particularly regarding the documented intermediate purchases and sales. The purchases may include imports, and the sales may also include sales for final consumptive use. Another source of error stems from the distributions and assumptions made for firm out-degree and in-degree distributions. While these assumptions are supported by stylized facts from Japanese firms (for out-degrees) and payment data from a large sample of Dutch firms (for in-degrees), it cannot be assumed that the parameters used in the reconstruction are universally applicable or representative of ‘true values'. Finally, the matching procedure results in a deterministic network where the ‘best' users have priority in connecting with their more closely aligned suppliers. This algorithm cannot account for noisy behavior or real-world uncertainties. In fact, for the 2012 version, with similar assumptions on degree distributions and the same matching algorithm, it has been demonstrated that these assumptions lead to biases in core network statistics such as the number of links in commodity layers~\cite{rachkov_potential_2021}, when compared with the ground-truth provided by a known sample of firm-to-firm connections collected by Dun \& Bradstreet (for 2012).
While aggregation at the SBI5 level is bound to reduce the biases that arose at the firm-level, it is still not clear how much the results are impacted by the propagation of these errors. Further discussion on limitations is provided at the beginning of the Discussion Section.

\subsection*{Network randomization methods}
The main goal of Network randomization methods is the generation of a statistical ensemble of networks, which are maximally random given available data.
In our case, we randomize each product layer of our industry-multilayer network separately using Maximum-Entropy methods. 
The available data - encoded as constraints in the Entropy maximization - consists of the supplier's(user's)  tendency to supply(use) a specific commodity and its output(input).
The obtained statistical ensemble of networks represents the possible realizations of the system taking into account suppliers' and users' tendencies.
After the generation of the synthetic ensemble of networks it is possible to extract metrics of interest as ensemble averages.

The null models we take into account are the Directed Binary Configuration Model (DBCM)~\cite{squartini_randomizing_2011} and the Reciprocal Binary Configuration Model (RBCM)~\cite{squartini_analytical_2011} for the estimation of network links, and the Conditional Reconstruction Method A (CReM$_{A}$)~\cite{parisi_faster_2020} and the newly developed Conditionally Reciprocal Weighted Configuration Model (CRWCM) for the conditional estimation of network weights.
The DBCM corresponds to the model that maximizes the Shannon Entropy attached to the distribution of  possible binary adjacency matrices, given that in-degree and out-degree  distributions are constrained on average.
The RBCM is also used for estimation of links by maximizing the Shannon Entropy attached to the distribution of  possible binary adjacency matrices, but makes use of additional information, namely the non-reciprocated out-degree, in-degree and the reciprocated degree distributions. These metrics are originated distinguishing links that are reciprocated from the ones that are not and summing on them.
Turning our attention to weighted networks, the CReM$_{A}$ is the Maximum-Entropy model that maximizes the conditional Shannon Entropy attached to the distribution of weighted networks, given the realization of the adjacency matrix $A$. The constraints used in the conditional optimization are the out-strength and in-strength distributions, corresponding to sum of weights going from and to a node, respectively.
The CRWCM, instead, is an augmented version of CReM$_{A}$, which can take better account of reciprocation by constraining the out-strength and in-strength distributions for reciprocated and non-reciprocated links.
Both CReM$_{A}$ and CRWCM are estimated using an annealed approach, following the articles~\cite{parisi_faster_2020, di_vece_deterministic_2023}, and consequently coupled with the relative binary model. Specifically directionality is encoded in the DBCM$+$CReM$_{A}$ model, also denoted as the \textit{directed} model, while directional and reciprocal information is encoded in the RBCM$+$CRWCM model, denoted as the \textit{reciprocated} model. For further information and the mathematical generation of link and weight distributions, please refer to the Method section.

\subsection*{Measuring empirical reciprocity statistics}

\begin{table*}[t!]
\centering
\begin{tabular}{c||c|c|c|c|c}
\hline
Layer-Statistics & Min & Lower Quartile & Median & Upper Quartile & Max \\
\hline
\hline
$N$ & $4$ & $62$ & $149$ & $544$ & $822$\\
$L$ & $3$ & $203$ & $678$ & $2076$ & $15198$\\
$W_{tot}$ & $0.95$ & $239$ & $768$ & $2027$ & $23767$ \\
\hline
\hline
$r_{t}$ & $0$ & $0.01$ & $0.05$ & $0.14$ & $0.78$\\
$r_{w}$ & $0$ &  $0.01$ & $0.08$ & $0.28$ & $0.78$\\
\hline
\hline
\end{tabular}
\caption{Description of the distribution of statistics such as the number of active industries $N$, the number of links $L$, the total weight $W_{tot}$, the topological reciprocity $r_{t}$ and the weighted reciprocity $r_{w}$ across commodity layers of the inter-industry network.}
\label{tab1}
\end{table*}

The presence of data on product granularity gives us the opportunity to study heterogeneity across commodity layers.
Let us consider in Table~\ref{tab1}
the number of layer-active industries $N$, the number of links $L$, the total weight $W_{tot}$, and reciprocity measures such as the \textit{topological reciprocity} $r_{t}$, defined as the ratio of reciprocated links to $L$, i.e.  
\begin{equation}
    r_{t} = \dfrac{L^{\leftrightarrow}}{L} = \dfrac{\sum_{i,j \neq i}a_{ij}^{\leftrightarrow}}{\sum_{i,j \neq i}a_{ij}}.  
\end{equation}
and its \textit{weighted} counterpart $r_{w}$, defined as the ratio of total weight on reciprocated links to $W$, i.e.
\begin{equation}
    r_{w} = \dfrac{W_{tot}^{\leftrightarrow}}{W_{tot}} = \dfrac{\sum_{i,j \neq i}w_{ij}^{\leftrightarrow, out}}{\sum_{i,j \neq i}w_{ij}}.  
\end{equation}

The median for $N$ is $149$, meaning that for around $50\%$ of commodity layers there are less than $149$ active industries (as suppliers or users). At the same time, $25\%$ of commodity layers have less than $62$ industries, and another $25\%$ have more than $544$ industries. Consequently, industries are specialized among a small number of business activities for half of the commodity groups but, a small, and not negligible, number of layers is characterized by a high number of active industries and hence of industry heterogeneity. Some examples are suppliers of plastic goods that are sold to users with heterogeneous specializations, for instance, Bread, Beer, Cereals, Fish, etc.  
Also the distributions regarding the number of commodity-specific links $L$ and the related total weight $W_{tot}$ have wide distributions, with a minimum with few digits, respectively $3$ and $0.95$ (in millions of euro), and a maximum in 5 digits, respectively $15198$ and $23767$, implying a high degree of heterogeneity in network structure across commodity layers.

Passing from the commodity global statistics to $r_{t}$ and $r_{w}$, we see a high degree of heterogeneity also in this case, namely a minimum value of $0$ stands for layers where no link is reciprocated, i.e. users and suppliers represent two distinct sets of nodes (bipartite graph). Instead, in the majority of the commodities (above $75\%$) there is a not-null reciprocity. In fact, the median is respectively $0.05$ and $0.08$. There is also the presence of a small number of commodities (below $10\%$) which are characterized by a large reciprocity, with a maximum of $0.78$ for both $r_t$ and $r_w$.

Reciprocity can arise for different reasons: (1) the aggregation from firms to industries or (2) the aggregation of products.
To mention the first case, consider two firms A and B in the industry $i$ and other two firms C and D in industry $j$.
Suppose firm A supplies to firm D, while firm C supplies to firm B, in the same commodity layer. Once the firms are aggregated in the related industries, a reciprocated link emerges between them, even if reciprocity is not present at the firm level.\\
The second case follows from the fact that if each commodity layer represents a unique product, that could be represented by the finest CPA product classification (with around $5000$ products), and we take into account only intermediate supply and use, it is not reasonable to think that firms are at the same time suppliers and users (of that specific product). Instead, in case of product aggregation, firms may be suppliers of a product inside that commodity group and also users of another product inside that same commodity group.

\begin{figure*}
\centering 
\includegraphics[clip, trim=0cm 19.9cm 0cm 0cm,width=\linewidth]{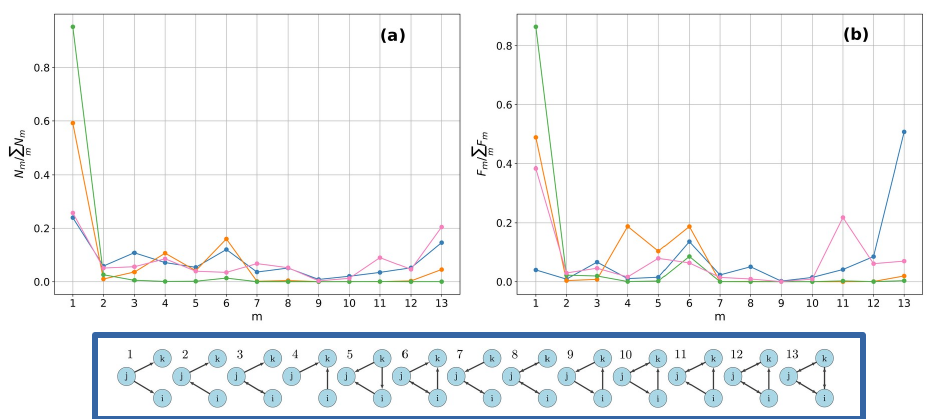}

\caption{Normalized Triadic Occurrences (a) and Fluxes (b): the Aggregated Network (\textcolor{blue}{$-\bullet-$}) presents a high occurrence of subgraphs $m=1$ and $m=13$, representing open-Vs and completely reciprocated triads, respectively. The latter covers most of the total amount of money traded. The Cereals  commodity layer (\textcolor{green}{$-\bullet-$}), with a high occurrence of subgraph $m=1$. A relatively high amount of money is distributed across $m=1$, $m=4$ and $m=6$. Gas/Hot Water/City Heating layer (\textcolor{orange}{$-\bullet-$}) with a predominant occurrence and flux in subgraph $m=1$. Agricultural Services layer (\textcolor{pink}{$-\bullet-$}), with a highly heterogenous spectrum of occurrences and fluxes. Completely cyclical triads have a high occurrence in the aggregated network, but break apart when passing to single commodity layers as G.H.C and Cereals, if not for rare cases such as Agricultural Services. In single commodity layers $m=1$ receives the highest concentration of money, signalling a large amount of money flows over structures that greatly depend on a limited number of suppliers, which control the market.}

\label{fig:3}
\end{figure*}

Let us now move to the analysis of triads.
We define \textit{triadic occurrences} $N_{m}$, the number of times a specific m-subgraph appears and \textit{triadic fluxes} $F_{m}$, the total amount of money circulating on each m-subgraph.
In Fig.~\ref{fig:3}, we depict their values normalizing by their sum across the $m$-types. The normalized $N_{m}$ and $F_{m}$ can be considered as the relative importance of a specific type of triadic subgraph in the network. The aggregated network (depicted in blue), where the weights of all commodity groups are summed, and three commodity layers, namely `Cereals' (in green), `Gas/Hot Water/City Heating (in orange) and `Agricultural Services' (in pink) are displayed.

In the aggregated network, the structures that occur relatively more are $m=1$, represented by a supplier connected to two users and $m=13$, the totally reciprocated cyclical triad. While $m=13$ is probably due to product aggregation, the predominance of $m=1$ is a signal of structural dependency on a limited number of suppliers. However, when normalized $F_{m}$ are investigated, $m=13$ still contain the majority of the volumes.
A similar profile, in the binary case, is given by the Agricultural Services, with the predominance of $m=1$ and $m=13$. At the same time a relatively smaller amount of money is concentrated on $m=13$ with respect to the aggregated case, while $m=1$ and $m=11$ carry a greater amount of money. During the product disaggregation weights on $m=13$ in the aggregated network are redistributed on other subgraphs, especially $m=1$.
In `Cereals' and `Gas/Hot Water/City Heating' these differences are even larger, with a relevant increase of triadic occurrences and fluxes on $m=1$, further increasing the dependency of the network on a limited amount of suppliers.
Note that when counting the different triads in Fig.~\ref{fig:3} they are not nested, i.e. a subgraph of type $m=8$ requires two reciprocated links and hence does not contain two subgraphs of type $m=1$, which contain only non-reciprocated links. Consequently, the number and fluxes over all triadic subgraphs are structurally independent across different types.

\subsection*{Binary Motif Analysis}

\begin{figure*}[ht!]
\includegraphics[clip, trim=0cm 4cm 0cm 1cm,width=0.8\linewidth,center]{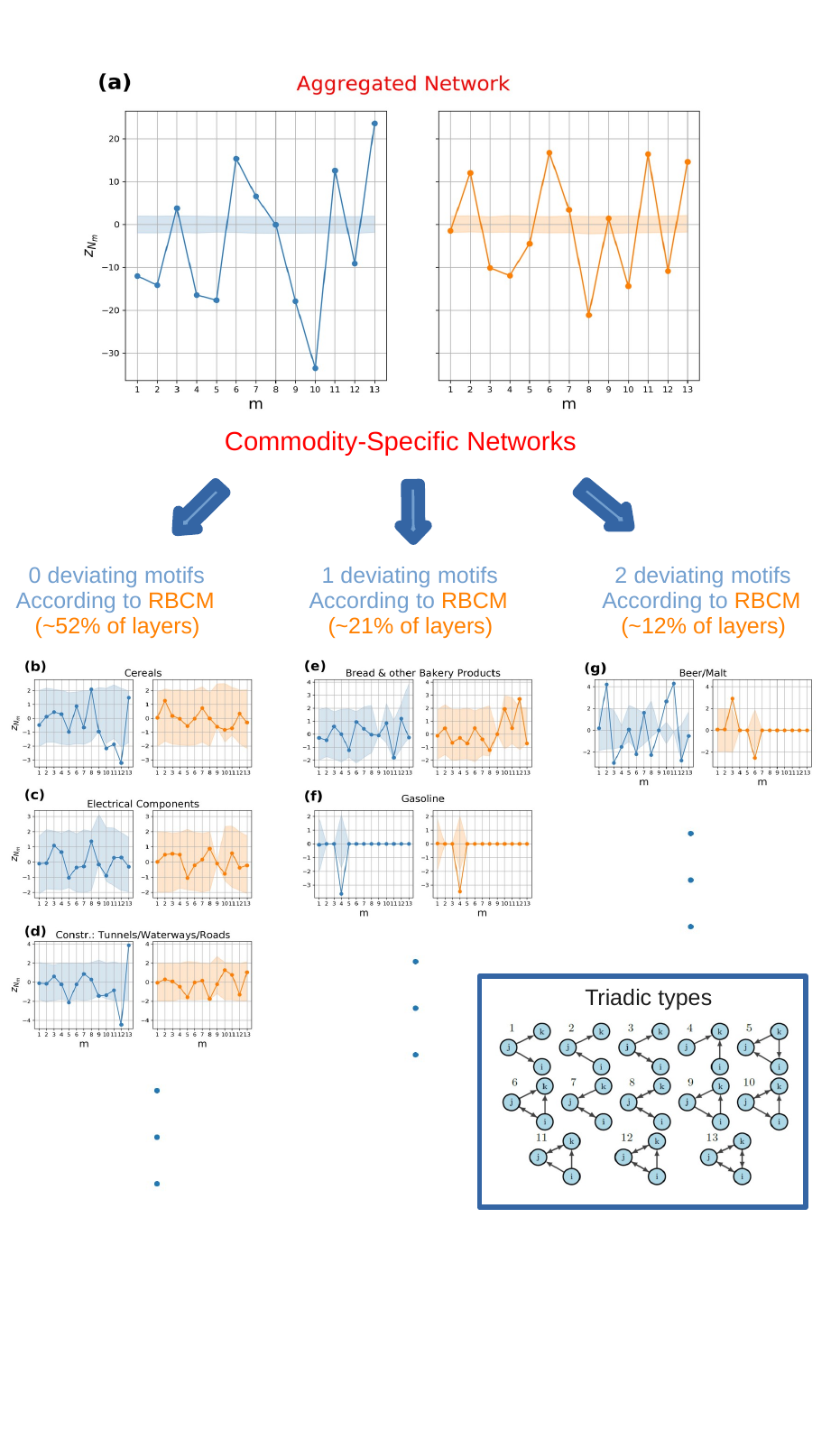}
\caption{Triadic binary motif analysis: DBCM (\textcolor{blue}{$\bullet$}) vs RBCM (\textcolor{orange}{$\bullet$}). (a) Analysis of the aggregated network with a single representative commodity. Numerous motifs and anti-motifs are present using DBCM and RBCM as null models. (b-d) Commodity groups where RBCM reproduces all the triadic structures, and they are, respectively, Cereals, Electrical Components, and the Construction of Tunnels, Waterways, and Roads. (e-f) Commodity groups with one network motif, namely Bread and Gasoline. (g) Commodity group with two network motifs, namely Beer/Malt. The CIs are computed by extracting the $2.5$-th and $97.5$-th percentile from an ensemble distribution of $500$ graphs. The numerous motifs and anti-motifs in the aggregated network can be seen as the aggregation of commodity groups presenting very few characteristic patterns.}\label{fig:4}
\end{figure*}

\begin{figure*}[ht!]
\centering

\includegraphics[width=\linewidth]{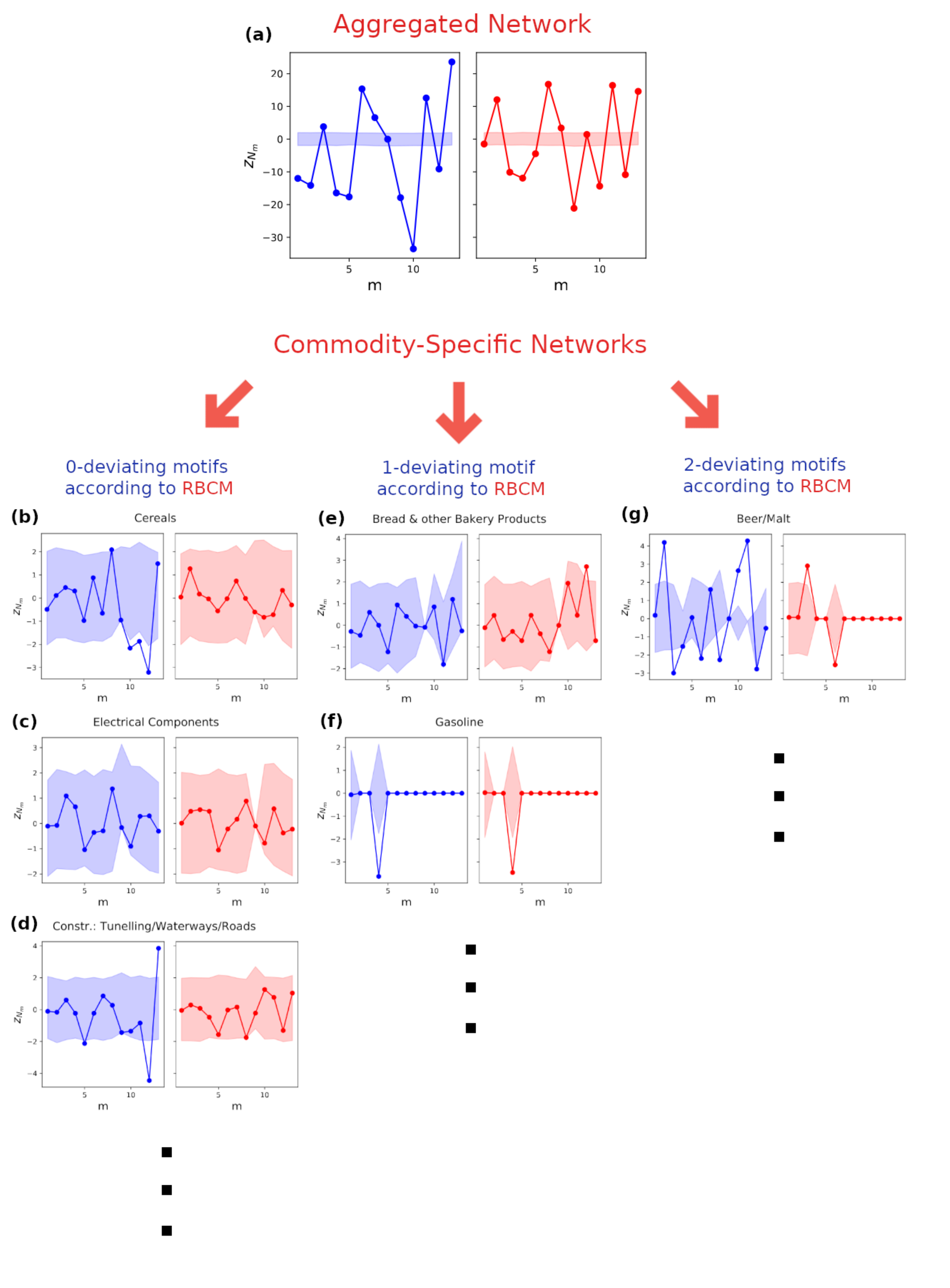}
\caption{Comparison DBCM (\textcolor{blue}{$\bullet$}) vs. RBCM (\textcolor{orange}{$\bullet$}): (a) Empirical Counter Cumulative Distribution Function $ECCDF$ of the number of deviating binary triadic motifs and anti-motifs across commodity layers. (b) Number of commodities $c_{h}(m)$ having a $m$-type motif (overoccurrence). (c) Number of commodities $c_{l}(m)$ having a $m$-type anti-motif (underoccurrence). RBCM explains more triadic structures than DBCM, as shown in the difference of their $ECCDF$. Passing from DBCM to RBCM reduces the number of $m$ motifs across commodities, with the exception of $m=6$, and anti-motifs, with the exception of $m=8$. The deviation of those triads is, hence, due to three-node correlations that go beyond directional and reciprocal tendencies of supply/use among industries. RBCM, hence, signals an increased vulnerability to demand shocks originating from the bankrupcty of industries of type $k$ in sub-types $m=6$ and an increased resiliency to supply/demand shocks of industries of type $j$ in triadic formations $m=8$.}
\label{fig:5}
\end{figure*}

We analyze the number of occurrences $N_{m}$ of all the possible triadic connected subgraphs, depicted in Fig.~\ref{fig1}(b). To quantify their deviations to randomized expectations, we define the \textit{binary z-score} of subgraph $m$ 
\begin{equation}\label{binary_z_score}
    z\left[N_{m}\right] = \dfrac{N_{m}(A^*) - \langle N_{m}\rangle}{\sigma\left[N_{m}\right]}
\end{equation}
where $N_{m}(A^*)$ is the number of occurrences of the $m$-type subgraph in the empirical adjacency matrix, $\langle N_{m} \rangle$ is its model-induced expected number of occurrences, and $\sigma \left[N_{m}\right]$ is the model-induced standard deviation.

An analytical procedure~\cite{squartini_analytical_2011} has been developed to compute the binary z-scores for the binary case. However, the assumption on the confidence intervals - represented as the interval $(-3,3)$ - holds true only if the ensemble distribution of $N_{m}$ is Normal for each $m$.
For all the commodities, $m$-types, and binary null models, we test the assumption using a Shapiro Test~\cite{shapiro_analysis_1965}. According to the test, $N_{m}$ ensemble distributions are in a large proportion not normal at the $5\%$ confidence level.
Consequently, we must use a numeric approach. Networks are sampled according to the DBCM recipe by (1) computing the induced connection probability $p_{ij; DBCM}$ and (2) establishing a link between industry $i$ and $j$ if and only if a uniformly distributed random number $u_{ij} \in U(0,1)$ is below $p_{ij; DBCM}$.
The analogous recipe for RBCM requires (1) computing the set of connection probabilities for non-reciprocated connection between $i$ and $j$, namely $p_{ij}^{\rightarrow}$, $p_{ij}^{\leftarrow}$ and $p_{ij}^{\not{\leftrightarrow}}$, and reciprocated connection $p_{ij}^{\leftrightarrow}$, generate a uniform random variable $u_{ij} \in (0,1)$ and (2) establishing the appropriate links in the dyad in the following way:
\begin{itemize}
    \item a non-reciprocated link from $i$ to $j$ if $u_{ij} \leq p_{ij}^{\rightarrow}$;
    \item a non-reciprocated link from $j$ to $i$ if $u_{ij} \in (p_{ij}^{\rightarrow},p_{ij}^{\rightarrow}+p_{ij}^{\leftarrow}]$;
    \item a reciprocated link from $i$ to $j$ (and from $j$ to $i$) if $u_{ij} \in (p_{ij}^{\rightarrow}+p_{ij}^{\leftarrow},p_{ij}^{\rightarrow}+p_{ij}^{\leftarrow}+p_{ij}^{\leftrightarrow}]$; 
    \item no links from $i$ to $j$ and from $j$ to $i$ otherwise.
\end{itemize}

In both cases, we generate a realization of $A$ and extract the $N_{m}$ statistic.
$\langle N_{m} \rangle$ and $\sigma\left[ N_{m}\right]$, are the average and standard deviation of $N_{m}$ extracted from the ensemble distribution of $500$ realizations of $A$.
After having computed $z\left[N_{m}\right]$, we also extract the $2.5$-th and $97.5$-th percentiles from the ensemble distribution of $N_{m}$ over all models and we standardize them using Eq.~(\ref{binary_z_score}) by replacing the empirical $N_{m}$ with the percentile. Such measures will serve as the $95\%$ CI for the z-score.

The results for the aggregated inter-industry network are in Fig.~\ref{fig:4}(a). The z-scores computed with respect to the DBCM are depicted in blue on the left panel, while the z-scores computed with respect to the RBCM are depicted in orange on the right panel. The corresponding confidence intervals at the $5\%$ percent are depicted with the same color (blue or orange) but in slight transparency.
The majority of $N_{m}$ are not reproduced by the randomized methods, i.e. the z-scores are outside the confidence intervals.
Specifically, only $N_{8}$ is reproduced by the DBCM, while both $N_{1}$ and $N_{9}$ are reproduced by the RBCM. Discounting reciprocal information does not only increase the number of triads that are statistically well described, but potentially changes their type, implying a \textit{qualitatively} different z-score profile. At the same time, in the aggregated picture, $m=1$ and $m=9$ are seen as described by a null model implementing reciprocity, i.e. neither high dependency on suppliers ($m=1$), nor unstable feedback loops ($m=9$), where industries supply to each other in a cyclical fashion, are revealed. The aggregated network, is hence, characterized by a multitude of structures that are not well described by the null model and are due to additional three-node correlations but is relatively resilient to supply shocks and cyclical input/output.
By disaggregating from the aggregated monolayer to the multi-commodity network, the majority of commodity-layers have triadic structures which are statistically reproduced by the reciprocal null model. 
Only $1$ or $2$ motifs or anti-motifs are present for the majority of the remaining commodities, a result indicating that beneath the aggregated picture, commodity groups are characterized by a small number of \textit{commodity-specific motifs} and \textit{anti-motifs}.

In Fig.~\ref{fig:4}(b-d) z-score profiles for three commodity layers are displayed, namely Cereals, Electrical Components, and the Construction of Tunnels, Waterways, and Roads. RBCM well describes all subgraph occurrences ($z_{N_{m}}$ is within CI), while the DBCM signals the presence of anti-motifs for $m=10$, $m=11$ and $m=12$ for Cereals, and anti-motif $m=12$ and motif $m=13$ for the Construction layer.
In Fig.~\ref{fig:4}(e-f) two z-score profiles are displayed - namely for Bread \& other Bakery Products and Gasoline - for which RBCM signals the presence of at least a motif or anti-motif. A motif $m=12$ is present for the former layer while an anti-motif for $m=4$ is present for the latter. Notice that for Bread the DBCM does not signal any motif or anti-motif, implying that deviations can emerge by introducing information on the reciprocal structure. Moreover, subgraph $m=9$ in Bread and the majority of subgraphs in the Gasoline commodity layer are characterized by a degenerate Confidence Interval: in all of the generated synthetic networks $N_{m=9}$ correspond to the empirical $N_{9}^*$ with null variance, i.e. the constraints imposed on the ensemble totally describe the specific $m$-type motif, a matter which can arise regardless of the lack of statistics in the related $N_{m}$. 
Finally, in Fig.~\ref{fig:4}(g) the z-profile for the commodity layer Beer/Malt is considered. The DBCM signals a large number of motifs, specifically for $m=2$, $m=10$, and $m=11$, and anti-motifs for $m=3$ and $m=8$. In contrast, the RBCM signals a lone motif $m=3$ and an anti-motif $m=6$.\\ 
In Fig.~\ref{fig:5}(a), the empirical counter cumulative distribution for the number of deviating binary triads is shown. Introducing reciprocal structure information reduces the number of motifs and anti-motifs present across commodities. For instance, the percentage of commodities with at least a motif or anti-motif is $61\%$ when compared to the DBCM, and $48\%$ when compared to 
 the RBCM, while the percentage of commodities having at least two motifs or anti-motifs is $46\%$ when compared to the DBCM and $27\%$ when compared to the RBCM.
 
Lastly, we identify the occurrence of $m$-type of motifs and anti-motifs across commodities by introducing two quantities, $c_{h}(m)$ and $c_{l}(m)$. $c_{h}(m)$ represents the number of commodities having a motif of type $m$  while $c_{l}(m)$ represents the same measure for anti-motifs.
The addition of the reciprocal structure reduces the number of commodity-specific motifs for each subgraph type, with the exception of motif $m=6$ as depicted in Fig.~\ref{fig:5}(b), and the number of anti-motifs for each type, with the exception of anti-motif $m=8$ as depicted in Fig.~\ref{fig:5}(c). The reciprocal null model, hence, reveals a higher number of commodities that are relatively more vulnerable to demand shock due to bankruptcy of industries of type $k$ in triadic formations $m=6$, while it reveals an increased resilience to supply/demand shocks originating from bankruptcy of industries of type $j$ in formations $m=8$.

\subsection*{Weighted Motif Analysis}

\begin{figure*}[ht!]
\includegraphics[clip, trim=0cm 4cm 0cm 0cm,width=0.75\textwidth,center]{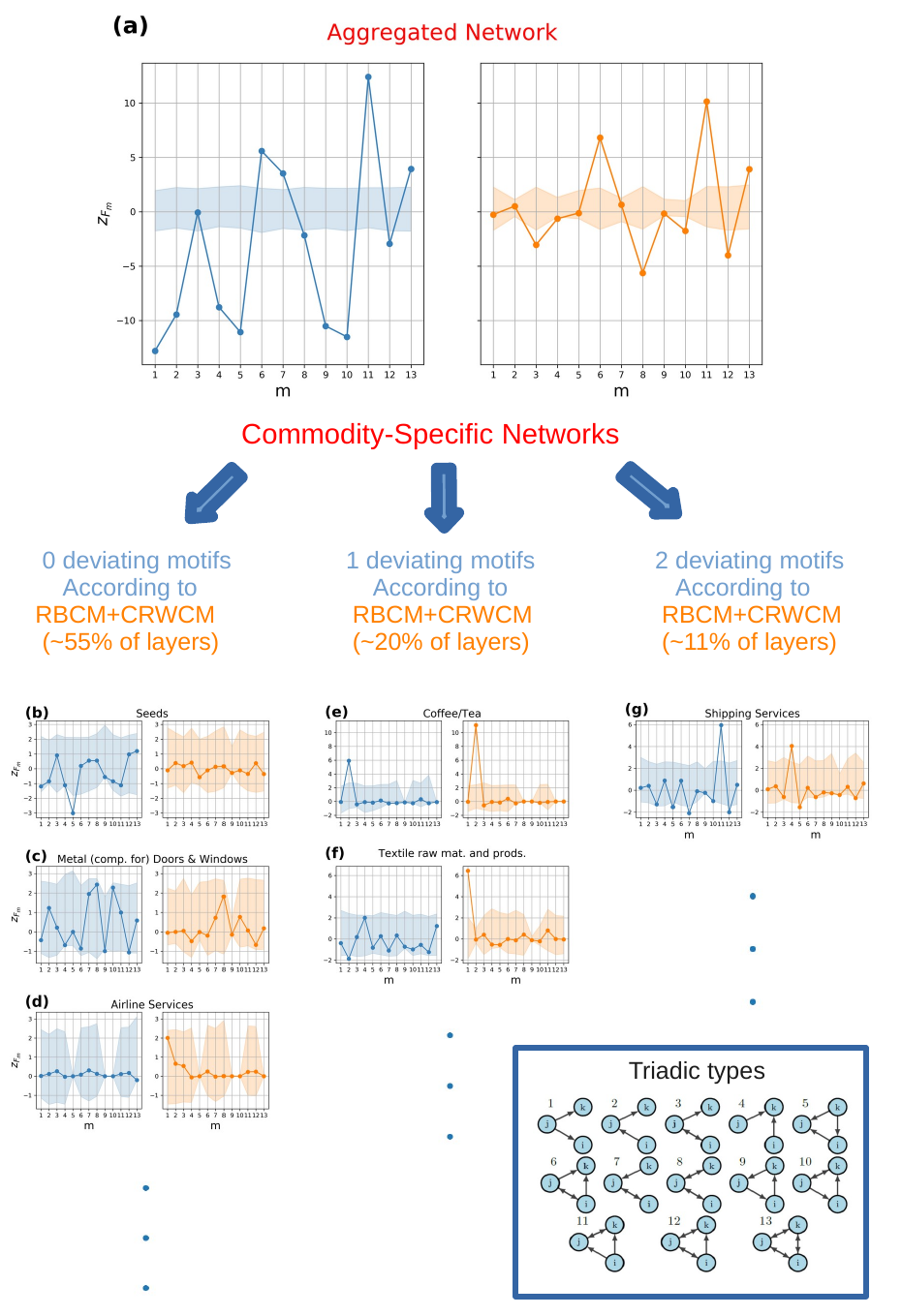}
\caption{
Triadic weighted motif analysis: DBCM+CReM$_{A}$ (\textcolor{blue}{$\bullet$}) vs RBCM+CRWCM (\textcolor{orange}{$\bullet$}). (a) Analysis of the aggregated network with a single representative commodity. A large number of motifs and anti-motifs are present when using DBCM+CReM$_{A}$, while three motifs are present when using the RBCM+CRWCM. (b-d) Commodity groups where RBCM+CRWCM reproduces all the triadic structures, and they are, respectively, Seeds, Metal Components for Doors \& Windows, and Airline Services. (e-f) Commodity groups with one network motif, namely Coffee/Tea and Textile raw materials and products. (g) Commodity group with two network motifs, namely Shipping Services. The CIs are computed by extracting the $2.5$-th and $97.5$-th percentile from an ensemble distribution of $500$ graphs. Passing from the aggregated network to the disaggregated product layers unveils the presence of a few commodity-specific motifs and anti-motifs.}
\label{fig:6}
\end{figure*}

\begin{figure*}[ht!]
\centering

\includegraphics[width=\linewidth,left]{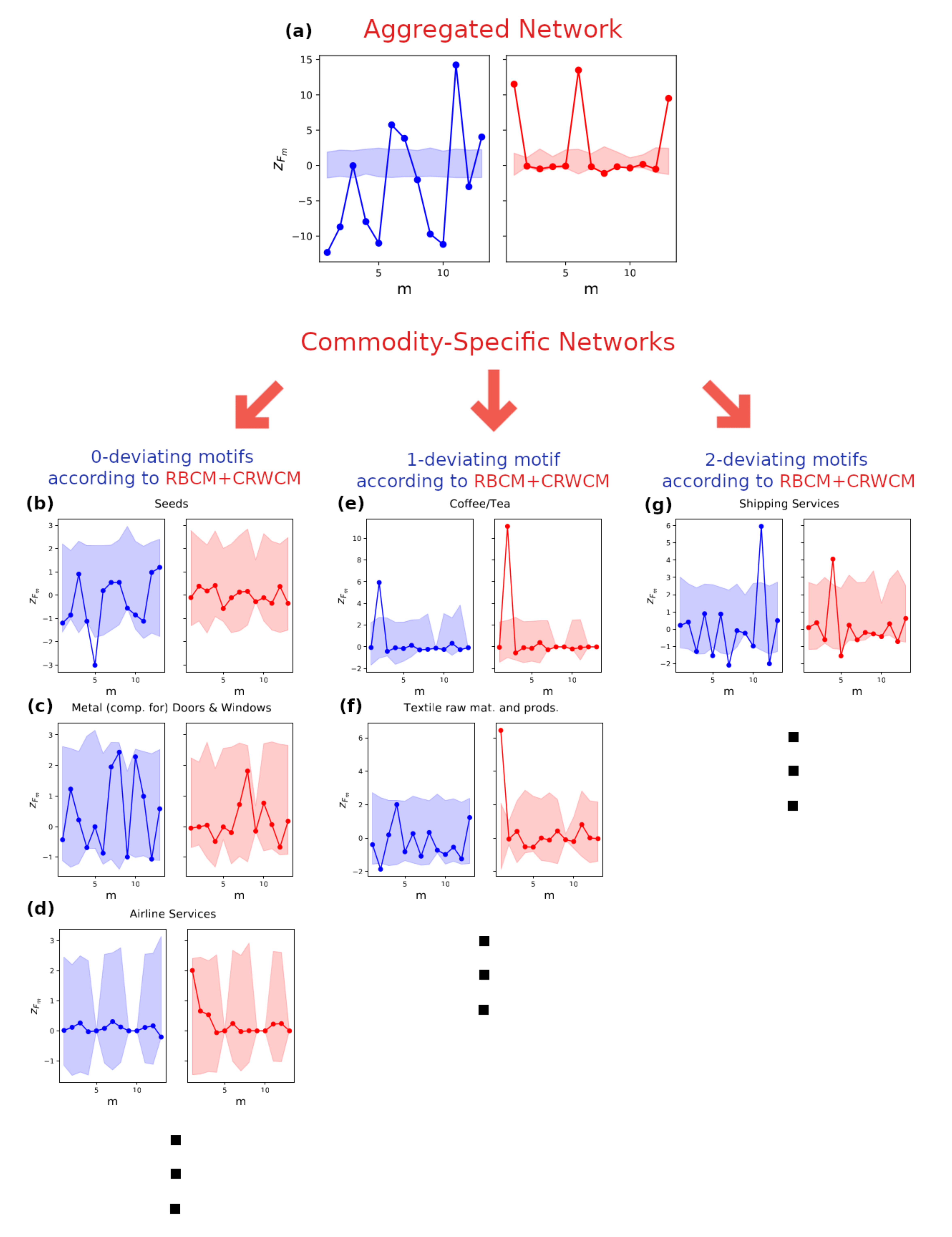}
\caption{
Comparison DBCM+CReM$_{A}$ (\textcolor{blue}{$\bullet$}) vs. RBCM+CRWCM (\textcolor{orange}{$\bullet$}): (a) Empirical Counter Cumulative Distribution Function $ECCDF$ of the number of deviating binary triadic motifs and anti-motifs across commodity layers. (b) The number of commodities $c_{h}(m)$ having a $m$-type motif. (c) The number of commodities $c_{l}(m)$ having a $m$-type anti-motif. RBCM+CRWCM explains slightly more triadic fluxes than DBCM+CReM$_{A}$, as shown in the difference of their $ECCDF$. Passing from the directed to the reciprocal model reduces the number of anti-motifs, with the exception of $m=8$. In contrast, it changes qualitatively the motif profile, with a slight dominance of $m=11$-type motifs when the directed model is used and a clear dominance of $m=1$-type motifs when the reciprocal model is used. The reciprocated model unveils a vulnerability to supply shocks originating from a decrease in supply volumes of industries of type $j$ in formations $m=1$.}
\label{fig:7}
\end{figure*}

While the bankruptcy of an entire industry is unrealistic, a shock due to a decrease in the flow of goods among industries can propagate along the supply chain, with side effects on the real economy. This implies that not only binary information is important for shock propagation but also weighted information, namely the amount of money circulating on connected structures.

Consider the \textit{triadic flux} $F_{m}$ on motif $m$, defined as the total money circulating on triadic subgraphs of type $m$. 
We characterize the deviation of empirical $F_{m}$ to null models by defining the \textit{weighted z-scores} as
\begin{equation}
    z\left[F_{m}\right] = \dfrac{F_{m}(W^*) - \langle F_{m}\rangle}{\sigma\left[F_{m}\right]}
\end{equation}
where $\langle F_{m} \rangle$ is the model-induced average amount of money circulating on motif $m$ and $\sigma\left[F_{m}\right]$ represents the model-induced standard deviation over the ensemble of network realizations.

The theoretical benchmark (or null model) is built by using a combination of binary and conditional weighted models, depending on the wanted constraints.
If we deem reciprocal information of negligible importance we should use the combination of models given by DBCM, for the sampling of the binary adjacency matrix, and the CReM$_{A}$, constraining the out-strength and in-strength sequences. 
If we deem reciprocal information necessary, a combination of the RBCM and CRWCM should be used. We compare here the two to establish the importance of the addition of reciprocity information for the detection of weighted motifs.

In operative terms, using a two-step model such as the DBCM+CReM$_{A}$ reduces to (1) establishing a link between industries $i$ and $j$ when a uniform random number $u_{ij} \in U(0,1)$ is such that $u_{ij} \leq p_{ij; DBCM}$, (2) if $i$ and $j$ are connected, sampling $w_{ij}$ by using the inverse transform sampling method technique, i.e., we generate a uniformly distributed random variable $\eta_{ij} \in U(0,1)$ such that 
\begin{equation}
F(v_{ij})=\int_0^{v_{ij}}q_{CReM_{A}}(w_{ij}|a_{ij}=1)dw_{ij} = \eta_{ij},
\end{equation}
then we invert the relationship finding the weight $v_{ij}$ to load on the link $(i,j)$.

The network sampling for the RBCM+CRWCM follows the same concepts with two major differences:
(1) a link is established using the RBCM recipe and 
(2) the dyadic conditional weight probability $q_{CReM_{A}}(w_{ij}|a_{ij}=1)$ is substituted with $q_{CRWCM}(w_{ij}|a_{ij}=1)$ in the inverse transform sampling.

In Fig.~\ref{fig:6}(a) the z-score profile for the aggregated network with a single representative commodity is depicted using the directed (in blue on the left panel) or the reciprocal models (in orange on the right panel).
There is a large number of motifs and anti-motifs when the benchmark model is directed, only $F_{3}$ does not deviate significantly. 

When reciprocity information is considered, the picture changes: only three motifs, namely $m=6$, $m=11$, and $m=13$, are identified, and four anti-motifs, namely $m=3$, $m=8$, $m=10$, and $m=12$, are found when the reciprocal null model is employed. This model's enhanced accuracy unveils a higher-than-expected volume of financial activity on sub-types characterized by a single exclusive user and two suppliers utilizing each other's products ($m=6$), two users supplying to each other while employing a product from the same supplier ($m=11$), and entirely cyclical triads ($m=13$). In contrast, a lower-than-expected level of financial activity transpires in open triads with two reciprocated ties ($m=8$), one reciprocated link and one exclusive user ($m=3$), or in closed triads of type $m=10$ and $m=12$. While it might be contended that the heightened concentration of funds on $m=13$ is attributable to aggregation bias, it is crucial to recognize that aggregation solely accounts for the increased monetary worth of the particular sub-type in absolute terms, not for the weighted motif obtained after adjusting for the statistical null model. It should be noticed that the emergence of these specific motifs cannot be easily explained without delving into greater detail, given the representative commodity scheme, while the picture cannot be merely reduced to a higher activity on open triads and a lower activity on closed triads.

Similarly to the binary case, passing from the aggregated network to the disaggregated product-level layers, it is possible to identify a small number of \textit{commodity-specific} weighted motifs and anti-motifs.

In Fig.~\ref{fig:6}(b-d) three commodity layers are depicted for which no motifs and anti-motifs are present when z-scores are computed using the reciprocal model. They are ‘Seeds', ‘Metal components for Doors \& Windows' and ‘Airline Services'.
In the ‘Seeds' layer, the directed model signals the presence of an anti-motif for $m=5$. In the second layer, no deviations are registered by both null models but CIs are of different nature, in fact, the reciprocal model allows a more restricted range of z-scores with respect to the directed model for $m=9$.
In the ‘Airline Services' layer, for both models, no deviations are present and three CIs are degenerate for $m=5$, $m=9$, and $m=10$.
 In Fig.~\ref{fig:6}(e-f) the z-scores relative to the commodity groups ‘Coffee/Tea' and ‘Textile raw materials and products' are depicted, for which 1 motif is present by using the reciprocal model.
 For both the directed and reciprocal models there is a weighted motif $m=2$ in the ‘Coffee/Tea' layer. In contrast, in the Textile products layer the directed model signals an anti-motif for $m=2$, while the reciprocal model signals a motif for $m=1$.
 If Fig.~\ref{fig:6}(g) the z-score profile for the commodity layer ‘Shipping Services' is shown: the directed model signals a large number of anti-motifs, specifically for $m=5$, $m=7$ and $m=12$, while it registers a motif for $m=11$.
 The reciprocal model, instead, registers a motif for $m=4$ and anti-motifs for $m=5$ and $m=12$. Different commodity layers call for different motifs and anti-motifs which are due to their specific characteristics. In this paper, we refrain from characterizing every single commodity layer, but a specific and thorough analysis is possible by visualizing the number of triadic sub-types, the z-score profile for $N_{m}$ and their weighted analogs.

The empirical counter cumulative distribution ECCDF($\#$ deviating W$\Delta)$ for the number of deviating weighted triads is depicted in Fig.~\ref{fig:7}(a). The number of deviating triadic fluxes is steadily lower using the reciprocal model. $F_{m}$ are maximally random for $49\%$ when the directed model benchmark is used and for $55\%$ according to the reciprocal model.  
The reduction of the number of motifs is however not as significant as in the binary case.

In Fig.~\ref{fig:7}(b-c) we plot the weighted analogous of $c_{h}(m)$ and $c_{l}(m)$. Reciprocal information decreases the occurrence of all types of anti-motifs across commodities, with the exception of $m=8$. Instead, the profile induced by $c_{h}(m)$ is \textit{significantly} different using the two null models. For instance, according to the directed model, $F_{1}$ is almost always well predicted, instead, it is the most occurring motif according to the reciprocal model. At the same time, reciprocity unveils the dependency of more than $40$ commodity layers on the supply of a limited amount of suppliers, which in this case control the market. In fact, the high presence of $m=1$ weighted motif signals the vulnerability of the industry-industry network to supply shocks provoked by a reduction of supply volumes.

\section*{Discussion}

The study of triadic motifs on production networks is still in its infancy due to a scarcity of reliable data. In the existing literature, only binary triadic motifs on one production network, the Japanese one, have been characterized for a single representative commodity~\cite{ohnishi_network_2010}, while the Hungarian dataset has been analyzed only for triadic occurrences without recurring to a null model~\cite{borsos_unfolding_2020}. The Japanese study revealed a simple but significant pattern: open triadic subgraphs are over-represented while closed triadic subgraphs are under-represented. This phenomenon was attributed to complementarity, where economic actors connect in tetradic structures - better explained by open triads - due to complementary needs~\cite{mattsson_functional_2021}.

Our findings corroborate the notion that an analysis based on a single representative commodity is insufficient to fully characterize a production network. Product-level data is \textit{essential} for disaggregating the network into layers that are characterized by commodity-specific binary motifs and anti-motifs. Moreover, we found that the majority of layers exhibit maximally random triadic structures when the reciprocal structure is considered.

At the level of binary motifs, we detected that cyclical reciprocated triadic subgraphs, which are dominant in the aggregated network, break up in the disaggregated product layers, where open triangles become dominant, especially $m=1$. However, using the RBCM as a benchmark, we proved that $m=1$ is always well described. Conversely, the completely cyclical triads, even if partially broken in the disaggregated layers, are often over-represented compared to the benchmark estimate. In general, constraining the reciprocation capacity of industries - by constraining the reciprocated degrees - is of the foremost importance when characterizing triadic motifs, as explained by the better accuracy and the decrease in binary triadic motifs and anti-motifs when using RBCM as a benchmark compared to DBCM.

We also characterized weighted motifs and anti-motifs, defined as the amount of money circulating on triadic subgraphs, with a novel model which constrains strengths, decomposing them according to the character of the corresponding links. This type of analysis is totally novel in the context of production networks, and rarely seen with benchmark models~\cite{picciolo_weighted_2022}. We find a non-trivial result already when analyzing the aggregated network, subgraphs that are well explained in binary terms - their occurrence is well described by the statistical ensemble induced by the DBCM or RBCM - can be not well described in weighted terms, meaning that even if a binary triadic subgraph has the expected occurrence it can accommodate an unexpected concentration of money. 
Furthermore, we identified a high presence of $m=1$ weighted motifs across commodity layers, a signal of commodity-specific dependency on a limited number of suppliers, which control the market. This implies that a large number of layers are vulnerable to supply shocks, which can arise due to a decrease in supplied volumes (and not only to the supplier's bankruptcy as in the binary case).

Changing the benchmark from a directed to a reciprocal model significantly changes the identity of motifs and anti-motifs across commodities.
Hence, it is essential to take into account the type of the corresponding link in which weights are sampled by constraining reciprocated and non-reciprocated strengths.

Overall, our results indicate that product-level information is \textit{strictly} necessary to identify triadic structures and fluxes in production networks. We hope that our study can encourage Statistics Bureaus around the world to implement policies and techniques to reveal or reconstruct a reliable product heterogeneity for firm-level transaction data.
Our analysis also shows that most commodity-specific layers can be reconstructed via null models that incorporate reciprocity while maintaining dyads independent. For these layers, network reconstruction methods of the type introduced in \cite{ialongo_reconstructing_2022}, if extended to incorporate reciprocity, are likely to perform well in replicating the properties of the entire layers starting from partial, node-specific information. 
Most other layers show at most one or a couple of deviating triadic motifs that are unexplained by the null model.
For these layers, additional information is needed to achieve a good reconstruction.
Once a rigorous product analysis has been performed, experts in the single commodity can interpret why such triadic formations over-occur or under-occur, accommodating an excessive or insufficient amount of trade volume, unveiling the detailed structure of the commodity-specific production networks.

In order to suggest improvements for further research, we conclude by noticing that our study is subject to two main limitations. First, an industry-level analysis inherently yields results that differ from those obtained from firm-level studies and underestimates the risk associated with exogenous and endogenous shocks~\cite{diem_estimating_2023}. 
Second, the dataset analyzed pertains to industries at the SBI5 level, a classification intermediate between firms and SBI4-level industries. Analyzing the dataset at the firm level was not feasible due to potential biases arising from the deterministic imputation and degree distribution assumptions, which are exacerbated when dealing with highly granular data. Conversely, analyzing industries at the SBI4 level, which encompasses a maximum of 132 industries, would imply that for a substantial number of commodities, very few industries are active. Consequently, the null model would trivially replicate, in a statistical sense, the triadic structures for the majority of commodity layers due to a lack of relevant observations. However, the same biases anticipated at the firm level can arise, even if mitigated, by selecting SBI5-level industries. This could potentially lead to biases in our analysis, especially in the type of motifs and anti-motifs found for each commodity. However, in order to validate all of our `fingerprints' we would need fully empirical data for industries at the SBI5 level for each of the $187$ commodities, an information that is not available in any country until now, to the best of our knowledge.

\section*{Methods}

\subsection*{Binary Null Models}\label{Sec_Binary_Null_Models}
For binary-directed graphs, the Maximum Entropy formalism prescribes the maximization of the Graph Entropy functional $S[P(A)]$
\begin{equation}
    S[P(A)] = - \sum_{A \in \mathbf{A}} P(A) \ln P(A)
\end{equation}
subject to the normalization of the Graph Probability $P(A)$ and to the constraints on network properties $C_{\alpha}^{*}$, i.e. 
\begin{equation}
\begin{cases}
    \sum_{A \in \mathbf{A}} P(A) &= 1 \\
    \sum_{A \in \mathbf{A}} P(A) C_{\alpha}(A) &= C_{\alpha}^{*}, \quad \forall \alpha,
\end{cases}
\end{equation}
hence maximizing the unbiasedness of the resulting $P(A)$ given available data.
Solving the optimization problem, we obtain the canonical $P(A)$
\begin{align}\label{P(A)_eq}
    P(A) &= \dfrac{e^{-\sum_{\alpha}\theta_{\alpha}C_{\alpha}^{*}(A)}}{\sum_{A \in \mathbf{A}}e^{-\sum_{\alpha}\theta_{\alpha}C_{\alpha}^{*}(A)}} = \nonumber \\
    &= \dfrac{e^{-H(A)}}{\sum_{A \in \mathbf{A}}e^{-H(A)}}
\end{align}
where $H(A)$ is denoted as the \textit{Graph Hamiltonian} and is defined as
\begin{equation}
    H(A) \equiv \sum_{\alpha} \theta_{\alpha} C_{\alpha}(A)^*.
\end{equation}

In this section, we focus on the binary reconstruction methods taking into account \textit{local} properties. 
\paragraph*{The Directed Binary Configuration Model}

In the \textit{Directed Binary Configuration Model} (DBCM), we choose as local properties the the out-degree $(k_i^{out})$ and the in-degree $(k_i^{in})$ representing the number of industries industry $i$ sells to and the number of industries industry $i$ buys from respectively. 

Out-degrees and in-degrees can be defined mathematically in terms of the adjacency matrix $A =(a_{ij})$ as
\begin{equation}
    \begin{cases}
    k_i^{out} &= \sum_{j\neq i} a_{ij} \\
    k_i^{in} &= \sum_{j\neq i} a_{ji}. 
    \end{cases}
\end{equation}
Solving the constrained Entropy maximization we obtain the Graph Probability $P(A)$ in Eq.~(\ref{P(A)_eq}) where
\begin{equation}
    H(A) = \sum_{i} \alpha_{i}^{out} k_i^{out} + \alpha_{i}^{in} k_i^{in}  . 
\end{equation}
The Graph Probability $P(A)$ can be re-written as the product of Bernoulli trials
\begin{equation}
    P(A) = \prod_{i,j \neq i} (p_{ij})^{a_{ij}}(1-p_{ij})^{1-a_{ij}}
\end{equation}
where $p_{ij} = P(a_{ij}=1)$ denotes the probability of connection of supplier $i$ with user $j$ and is equal to
\begin{equation}
    p_{ij} = \dfrac{x_{i}^{out}x_{j}^{in}}{1+x_{i}^{out}x_{j}^{in}}
\end{equation}
with $x_{i}^{out} \equiv  e^{-\alpha_{i}^{out}}$ and $x_{i}^{in} \equiv  e^{-\alpha_{i}^{in}}$.
By Maximum Log-Likelihood Estimation (MLE) on the log-likelihood $\mathcal{L}=\ln(P(A))$ we obtain the Lagrange parameters $\alpha_{i}^{out}$ and $\alpha_{i}^{in}$ $\forall i$, a procedure equivalent to solving a system of $2N$ coupled equations
\begin{equation}
    \begin{cases}
        k_{i}^{out,*} &= \langle k_{i}^{out} \rangle = \sum_{j \neq i} p_{ij} \\
        k_{i}^{in,*} &= \langle k_{i}^{in} \rangle = \sum_{j \neq i} p_{ji}.
    \end{cases}
\end{equation}
where $N$ is the number of industries in the network and $\langle k_i ^{out} \rangle$ and $\langle k_i ^{in} \rangle$ denote the ensemble averages of out-degrees and in-degrees respectively.

\paragraph*{The Reciprocal Binary Configuration Model}

In the \textit{Reciprocal Binary Configuration Model} (RBCM), we decompose the degree according to the reciprocal nature of the connection at hand, namely in 
non-reciprocated out-degree $k_i^{\rightarrow}$,
non-reciprocated in-degree $k_i^{\leftarrow}$
and reciprocated degree $k_i^{\leftrightarrow}$.
Those measures can be defined mathematically in terms of the adjacency matrix $A =(a_{ij})$ as
\begin{equation}
\begin{cases}
    k_{i}^{\rightarrow} &= \sum_{j \neq i} a_{ij}(1-a_{ji}) = \sum_{j \neq i} a_{ij}^{\rightarrow}  \\
    k_{i}^{\leftarrow} &= \sum_{j \neq i} a_{ji}(1-a_{ij}) = \sum_{j \neq i} a_{ij}^{\leftarrow}  \\
    k_{i}^{\leftrightarrow} &= \sum_{j \neq i} a_{ij}a_{ji} = \sum_{j \neq i} a_{ij}^{\leftrightarrow}.
\end{cases}
\end{equation}
Solving the Constrained Maximization Entropy problem, we obtain the Graph Probability $P(A)$ as in Eq.~(\ref{P(A)_eq}) with Graph Hamiltonian given by
\begin{equation}
    H(A) = \sum_{i} \alpha_i^{\rightarrow} k_i^{\rightarrow} + \alpha_i^{\leftarrow} k_i^{\leftarrow} + \alpha_i^{\leftrightarrow} k_i^{\leftrightarrow}.
\end{equation}
The model-induced Graph Probability $P(A)$ is the product of Bernoulli trials of mutually exclusive events
\begin{equation}
    P(A) = \prod_{j < i} \left(p_{ij}^{\rightarrow}\right)^{a_{ij}^{\rightarrow}} \left(p_{ij}^{\leftarrow}\right)^{a_{ij}^{\leftarrow}}
    \left(p_{ij}^{\leftrightarrow}\right)^{a_{ij}^{\leftrightarrow}}
    \left(p_{ij}^{\nleftrightarrow}\right)^{a_{ij}^{\nleftrightarrow}}
\end{equation}
with 
\begin{equation}\label{RBCM_pijs}
    \begin{cases}
    p_{ij}^{\rightarrow} &= \dfrac{x_i^{\rightarrow}x_{j}^{\leftarrow}}{1+x_{i}^{\rightarrow}x_{j}^{\leftarrow}+x_{i}^{\leftarrow}x_{j}^{\rightarrow}+x_{i}^{\leftrightarrow}x_{j}^{\leftrightarrow}} \\
    p_{ij}^{\leftarrow} &= \dfrac{x_i^{\leftarrow}x_{j}^{\rightarrow}}{1+x_{i}^{\rightarrow}x_{j}^{\leftarrow}+x_{i}^{\leftarrow}x_{j}^{\rightarrow}+x_{i}^{\leftrightarrow}x_{j}^{\leftrightarrow}} \\
    p_{ij}^{\leftrightarrow} &= \dfrac{x_i^{\leftrightarrow}x_{j}^{\leftrightarrow}}{1+x_{i}^{\rightarrow}x_{j}^{\leftarrow}+x_{i}^{\leftarrow}x_{j}^{\rightarrow}+x_{i}^{\leftrightarrow}x_{j}^{\leftrightarrow}} \\
    p_{ij}^{\nleftrightarrow} &= \left[1+x_{i}^{\rightarrow}x_{j}^{\leftarrow}+x_{i}^{\leftarrow}x_{j}^{\rightarrow}+x_{i}^{\leftrightarrow}x_{j}^{\leftrightarrow}\right]^{-1}.
    \end{cases}
\end{equation}
where $x_{i}^{\rightarrow} \equiv e^{-\alpha_{i}^{\rightarrow}}$, $x_{i}^{\leftarrow} \equiv e^{-\alpha_{i}^{\leftarrow}}$ and $x_{i}^{\leftrightarrow} \equiv e^{-\alpha_{i}^{\leftrightarrow}}$ are the exponentiated Lagrange multipliers tuning for the non-reciprocated out-degree, non-reciprocated in-degree and reciprocated degree respectively.
The Lagrange multipliers $\alpha_{i}^{\rightarrow}$, $\alpha_{i}^{\leftarrow}$ and $\alpha_{i}^{\leftrightarrow}$
are found using MLE on the Log-likelihood $\mathcal{L} = \ln(P(A))$, a procedure equivalent to solving the system of $3N$ coupled equations reading

\begin{equation}
    \begin{cases}
        k_{i}^{\rightarrow} &= \langle k_{i}^{\rightarrow} \rangle = \sum_{j \neq i} p_{ij}^{\rightarrow}\\
        k_{i}^{\leftarrow} &= \langle k_{i}^{\leftarrow} \rangle = \sum_{j \neq i} p_{ij}^{\leftarrow}\\
        k_{i}^{\leftrightarrow} &= \langle k_{i}^{\leftrightarrow} \rangle = \sum_{j \neq i} p_{ij}^{\leftrightarrow},
    \end{cases}
\end{equation}
i.e., equating the reciprocated and non-reciprocated degrees to their ensemble averages.

\subsection*{Conditional Weighted Null Models}\label{Sec_Conditional_Weighted_Null_Models}
When inspecting network weights, the numeric character of the involved trade volumes restricts the basket of available models. If the weights are discrete-valued, the constrained Entropy maximization leads to a family of geometric distributions~\cite{di_vece_gravity_2022, squartini_randomizing_2011-1,squartini_reciprocity_2013}. In contrast, continuous values lead to a family of exponential probability distributions when the constraints arise from node-specific properties~\cite{parisi_faster_2020, di_vece_reconciling_2023}.
We treat the conditional problem, which is well defined only after deciding the form of the binary adjacency matrix $A$.

The conditional Graph Entropy $S[Q(W|A)]$, measuring the uncertainty attached to the probability of having a weighted adjacency matrix $W$ compatible with a given realization of the binary adjacency matrix $A$, i.e.
\begin{equation}
    S[Q(W|A)] = - \sum_{A \in \mathbf{A}} P(A) \int_{W_{A}} Q(W|A) \ln Q(W|A) dW
\end{equation}
is maximized given the normalization of the conditional weighted probability density function $Q(W|A)$ and the constraints $C_{\alpha}(W)$  
\begin{equation}
\begin{cases}
    \int_{W_{A}} Q(W|A)dW &= 1 \\
    \sum_{A} P(A) \int_{W_{A}} Q(W|A) C_{\alpha}(W) dW &= C_{\alpha}^*, \quad \forall \alpha.
    \end{cases}
\end{equation}
where the set of $C_{\alpha}^*$ represent known node-specific properties.
From this constrained conditional maximization we obtain $Q(W|A)$, as
\begin{equation}\label{Q(W|A)_eq}
    Q(W|A) = \begin{cases}
     \dfrac{e^{-H(W)}}{\int_{W_{A}}e^{-H(W)} dW_{A}} \quad &W \in W_A \\
     0 &W \notin W_A
     \end{cases}
\end{equation}
where $W_A$ stands for the ensemble of realizations of $W$ compatible with $A$ (with weights sampled only on connected dyads $a_{ij}=1$) and the Graph Hamiltonian $H(W)$ is defined as 
\begin{equation}
    H(W) \equiv \sum_{\alpha} \beta_{\alpha} C_{\alpha}(W).
\end{equation}

Parameters $\beta_{\alpha}$ are estimated using MLE on the log-likelihood function $\mathcal{L}_{W}$ reading
\begin{equation}
    \mathcal{L}_{W|A} = - H_{\Vec{\beta}}(W) - \ln(Z_{\Vec{\beta},A}) 
\end{equation}
where $Z_{\Vec{\beta},A}$ is the \textit{conditional partition function} and its computation is possible only if total information about $A$ is available.
However, estimating parameters on the empirical topology $A$ neglects its intrinsic random variability when it is sampled using a binary model, such as DBCM or RBCM. This problem is solved in Network Science literature by defining
the \textit{generalized log-likelihood} $\mathcal{G_{\Vec{\beta}}}$~\cite{parisi_faster_2020, di_vece_deterministic_2023}
\begin{equation}
    \mathcal{G_{\Vec{\beta}}} = - H_{\Vec{\beta}}(\langle W \rangle) - \sum_{A \in \mathbf{A}} P(A) \ln(Z_{\Vec{\beta},A})
\end{equation}
where $P(A)$ is the Graph Probability induced by the binary model.
In the following, we mainly deploy the estimation based on $\mathcal{G_{\Vec{\beta}}}$ for weighted models.
Using the framework mentioned above, we can solve the conditional maximum Entropy problem taking into account \textit{weighted} local properties. 

\paragraph*{The CReM$_A$}

When randomizing the weighted adjacency matrix $W$, trade marginals such as the out-strength $s_{i}^{out}$ and the in-strength $s_{i}^{in}$ - representing the total output or total input of industry $i$ - are usually constrained~\cite{squartini_randomizing_2011-1,mastrandrea_enhanced_2014}.
The out-strength $s_i^{out}$ and the in-strength $s_i^{in}$ sequences are defined as the marginals of the weighted adjacency matrix $W$, namely
\begin{equation}
    \begin{cases}
    s_i^{out} &= \sum_{j\neq i} w_{ij} \\
    s_i^{in} &= \sum_{j\neq i} w_{ji}.
    \end{cases}
\end{equation}
Solving the constrained conditional Entropy maximization leads to a conditional cumulative function $Q(W|A)$ as in Eq.~(\ref{Q(W|A)_eq}) where
\begin{equation}
    H(W) = \sum_{i} \beta_i^{out} s_i^{out} + \beta_i^{in} s_i^{in}
\end{equation}
with a conditional Graph distribution
\begin{align}
    Q(W|A) &= \prod_{i,j \neq i; a_{ij}=1} q_{ij}(w|a=1) = \nonumber \\
    &= \prod_{i,j \neq i; a_{ij}=1} \left[\left(\beta_i^{out} + \beta_{j}^{in} \right) e^{-(\beta_i^{out} + \beta_{j}^{in})w_{ij}}\right]^{a_{ij}}
\end{align}
i.e. the product of dyadic exponential distributions in $w_{ij}$ conditional on the establishment of the link $a_{ij}$ and regulated by the node-specific Lagrange parameters $\beta_{i}^{out}$ and $\beta_{i}^{in}$ $\forall i$.
By using Generalized Log-likelihood Estimation (GLE), we find the Lagrange parameters - a procedure that equates to slightly changing the dyadic conditional probability by substituting $a_{ij}$ with a dyadic term $f_{ij}$ such that $f_{ij}= \langle a_{ij} \rangle$, i.e., $f_{ij}$ is the ensemble average of $a_{ij}$ and
\begin{equation}
    q_{ij}(w_{ij}|a_{ij}=1) = \left[ \left( \beta_{i}^{out} + \beta_{j}^{in} \right) e^{-(\beta_{i}^{out}+\beta_{j}^{in})} \right]^{f_{ij}}.
\end{equation}

Maximizing $G_{\Vec{\beta}}$ we obtain a system of $2N$ coupled equations reading 
\begin{equation}
    \begin{cases}
    s_i^{out} &= 
    \sum_{j\neq i} \dfrac{f_{ij}}{\beta_{i}^{out}+\beta_{j}^{in}} = \langle s_{i}^{out} \rangle \\
    s_i^{in} &= \sum_{j\neq i} \dfrac{f_{ji}}{\beta_{i}^{in}+\beta_{j}^{out}} = \langle s_{i}^{in} \rangle
    \end{cases}
\end{equation}
and find $\{\beta_i^{in},\beta_{i}^{out}\}$ for each industry.

\paragraph*{The CRWCM model}
In order to take into account reciprocity, we develop a novel model denoted as \textit{Conditionally Reciprocal Weighted Configuration Model} (CRWCM), that considers the different nature of links on which weights are sampled, namely reciprocated and non-reciprocated links.
This choice leads to the definition of four trade marginals for each supplier/user, namely
\begin{itemize}
    \item the non-reciprocated out-strength $s_{i}^{\rightarrow}$ which measures the output of supplier $i$ to users from which it does not buy, defined in terms of $W$ as
    \begin{equation}
        s_{i}^{\rightarrow} = \sum_{j \neq i} a_{ij}^{\rightarrow} w_{ij} = \sum_{j \neq i} w_{ij}^{\rightarrow}
    \end{equation}
    \item the non-reciprocated in-strength $s_{i}^{\leftarrow}$, which measures the input of industry $i$ from suppliers to which it does not itself supply, defined as
    \begin{equation}
        s_{i}^{\leftarrow} = \sum_{j \neq i} a_{ij}^{\leftarrow} w_{ji} = \sum_{j \neq i} w_{ij}^{\leftarrow}
    \end{equation}
    \item the reciprocated out-strength $s_{i}^{\leftrightarrow, out}$, measuring the output of supplier $i$ to users from which it also purchases, reading
    \begin{equation}
        s_{i}^{\leftrightarrow, out} = \sum_{j \neq i}  a_{ij}^{\leftrightarrow} w_{ij} = \sum_{j \neq i} w_{ij}^{\leftrightarrow, out}
    \end{equation}
    \item and the reciprocated in-strength $s_{i}^{\leftrightarrow, in}$, measuring the input of user $i$ from suppliers to which it also supplies, defined as 
    \begin{equation}
    s_{i}^{\leftrightarrow, in} = \sum_{j \neq i} a_{ij}^{\leftrightarrow} w_{ji} = \sum_{j \neq i} w_{ij}^{\leftrightarrow, in}
    \end{equation}
\end{itemize}

Solving the constrained conditional Maximum Entropy problem, we obtain the Conditional Weighted Graph Probability in Eq.~(\ref{Q(W|A)_eq}) where the Graph Hamiltonian is given by
\begin{equation}
    H(W) = \sum_{i}  \beta_{i}^{\rightarrow} s_{i}^{\rightarrow} + \beta_{i}^{\leftarrow} s_{i}^{\leftarrow} + \beta_{i}^{\leftrightarrow, out} s_{i}^{\leftrightarrow, out} + \beta_{i}^{\leftrightarrow, in} s_{i}^{\leftrightarrow, in} 
\end{equation}
leading to 
\begin{align}
Q(W|A) &= \prod_{j \neq i, a_{ij}=1} q_{ij}(w|a_{ij}=1) = \nonumber \\ 
\end{align}
where $q_{ij}(w|a_{ij})$ for the single dyad depends on the possible states of $w_{ij}$, namely
\begin{equation}
    \begin{cases}
(\beta_{i}^{\rightarrow}+\beta_{j}^{\leftarrow})e^{-(\beta_{i}^{\rightarrow}+\beta_{j}^{\leftarrow})w_{ij}^{\rightarrow}}  &\text{for $w_{ij}^{\rightarrow} > 0$} \\
(\beta_{i}^{\leftrightarrow,out}+\beta_{j}^{\leftrightarrow,in})e^{-(\beta_{i}^{\leftrightarrow,out}+\beta_{j}^{\leftrightarrow,in})w_{ij}^{\leftrightarrow, out}} &\text{for $w_{ij}^{\leftrightarrow, out} > 0$} \\
        0 &\text{for $w_{ij} = 0$}.
    \end{cases}
\end{equation}

Rephrasing the vector $\{ a_{ij}^{\rightarrow},a_{ij}^{\leftarrow},a_{ij}^{\leftrightarrow},a_{ij}^{\not{\leftrightarrow}}\}$ of $a_{ij}$-states
into the vector of their ensemble averages $\{ f_{ij}^{\rightarrow},f_{ij}^{\leftarrow},f_{ij}^{\leftrightarrow},f_{ij}^{\not{\leftrightarrow}}\}$, where $f_{ij}^{(\cdot)} = \langle a_{ij}^{(\cdot)} \rangle$ depends on the binary model of choice, we can use GLE for the estimation of the $4N$ parameters. The resulting generalized log-likelihood is separable in a reciprocal and non-reciprocal component, i.e., $\mathcal{G}_{\Vec{\beta}} = \mathcal{G}_{\Vec{\beta}}^{\rightarrow} + \mathcal{G}_{\Vec{\beta}}^{\leftrightarrow}$ (see Appendix B for details).
The Lagrange parameters $\Vec{\beta}$ are retrieved by maximizing $\mathcal{G}_{\Vec{\beta}}$, which equates to solving two uncoupled systems of $2N$ coupled equations reading
\begin{equation}
    \begin{cases}
    s_i^{\rightarrow} &= 
    \sum_{j\neq i} \dfrac{f_{ij}^{\rightarrow}}{\beta_{i}^{\rightarrow}+\beta_{j}^{\leftarrow}} = \langle s_{i}^{\rightarrow} \rangle \\
    s_i^{\leftarrow} &= \sum_{j\neq i} \dfrac{f_{ij}^{\leftarrow}}{\beta_{i}^{\leftarrow}+\beta_{j}^{\rightarrow}} = \langle s_{i}^{\leftarrow} \rangle
    \end{cases}
\end{equation}
for the non-reciprocated sub-problem and
\begin{equation}
    \begin{cases}
    s_i^{\leftrightarrow, out} &= 
    \sum_{j\neq i} \dfrac{f_{ij}^{\leftrightarrow}}{\beta_{i}^{\leftrightarrow, out}+\beta_{j}^{\leftrightarrow, in}} = \langle s_{i}^{\leftrightarrow, out} \rangle \\
    s_i^{\leftrightarrow, in} &= \sum_{j\neq i} \dfrac{f_{ij}^{\leftrightarrow}}{\beta_{i}^{\leftrightarrow, in}+\beta_{j}^{\leftrightarrow, out}} = \langle s_{i}^{\leftrightarrow, in} \rangle 
    \end{cases}
\end{equation}
for the reciprocated sub-problem.

\subsection*{Data Availability}
The data analyzed in this study is under licence by Statistics Netherlands (CBS). Requests to access data should be directed to FPP, f.pijpers@cbs.nl.

\subsection*{Code Availability}
The code is available as a Python package named `NuMeTriS - Null Models for Triadic Structures' and containing solvers and routines for triadic motif analysis for the mentioned models, namely the DBCM, the RBCM and the mixture models DBCM+CReM$_{A}$ and RBCM+CRWCM. 
The package is available at the following URL: \href{https://github.com/MarsMDK/NuMeTriS}{\texttt{https://github.com/MarsMDK/NuMeTriS}}.

\hspace{2em}

\subsection*{Acknowledgements}
SoBigData.it receives funding from European Union – NextGenerationEU – National Recovery and Resilience Plan (Piano Nazionale di Ripresa e Resilienza, PNRR) – Project: “SoBigData.it – Strengthening the Italian RI for Social Mining and Big Data Analytics” – Prot. IR0000013 – Avviso n. 3264 del 28/12/2021. This work has been also supported by the project ‘Network analysis of economic and financial resilience’, Italian DM n. 289, 25-03-2021 (PRO3 Scuole) CUP D67G22000130001 and by the PNRR-M4C2-Investimento 1.3, Partenariato Esteso PE00000013-“FAIR-Future Artificial Intelligence
Research”-Spoke 1 “Human-centered AI”, funded by the European
Commission under the NextGeneration EU programme. DG acknowledges support from the Dutch Econophysics Foundation (Stichting Econophysics, Leiden, the Netherlands)
and the Netherlands Organization for Scientific Research (NWO/OCW).
MDV and DG acknowledge support from the ‘Programma di Attività Integrata’ (PAI) project ‘Prosociality, Cognition and Peer Effects’ (Pro.Co.P.E.), funded by IMT School for Advanced Studies Lucca. MDV also acknowledges
support by the European Community programme under the
funding schemes: ERC-2018-ADG G.A. 834756 “XAI: Science and technology for the eXplanation of AI decision
making.”

\subsection*{Author Contributions}
Conceptualization, Methodology, and writing - review \& editing: all
authors; software, visualization, and writing –
original draft preparation: MDV; Supervision: FPP and DG; Data Access: FPP, Fund Acquisition: DG.


\begin{thebibliography}{99}
\section*{References}

\bibitem{acemoglu_network_2012} D.~Acemoglu, V.~M.~Carvalho, A.~Ozdaglar, and A.Tahbaz-Salehi, \emph{Econometrica} \textbf{80}, 1977 (2012).

\bibitem{aobdia_inter-industry_2014} D.~Aobdia, J.~Caskey, and N.~B.~Ozel, \emph{Review of Accounting Studies} \textbf{19}, 1191 (2014).

\bibitem{atalay_how_2017} E.~Atalay, \emph{American Economic Journal: Macroeconomics} \textbf{9}, 254 (2017).

\bibitem{bouakez_transmission_2009} H.~Bouakez, E.~Cardia, and F.~J.~Ruge-Murcia, \emph{International Economic Review} \textbf{50}, 1243 (2009).

\bibitem{brintrup_predicting_2018} A.~Brintrup, P.~Wichmann, P.~Woodall, D.~McFarlane,
E.~Nicks, and W.~Krechel, \emph{Complexity} \textbf{2018}, e9104387 (2018).

\bibitem{pichler_simultaneous_2022} A.~Pichler and J.~D.~Farmer, \emph{Economic Systems Research}
\textbf{34}, 273 (2022).

\bibitem{bacilieri_firm_2023} A.~Bacilieri, A.~Borsos, P.~Astudillo-Estévez and F.~Lafond, INET Oxford Working Paper No. 2023-08. (2023)

\bibitem{atalay_network_2011} E.~Atalay, A.~Hortaçsu, J.~Roberts, and C.~Syverson, \emph{Proceedings of the National Academy of Sciences} \textbf{108}, 5199 (2011).

\bibitem{bernard_production_2019} A.~B.~Bernard, A.~Moxnes, and Y.~U.~Saito, \emph{Journal of
Political Economy} \textbf{127}, 639 (2019).


\bibitem{buiten_reconstruction_2021}
G.~Buiten, E.~de Jong, G.~Mooijen, S.~Hooijmaaijers, P.~Bogaart, CBS Technical Paper, doi: 10.13140/RG.2.2.16685.77286 (2021). 


\bibitem{carvalho_supply_nodate} V.~M.~Carvalho, M.~Nirei, Y.~U.~Saito, and A.~Tahbaz-
Salehi, \emph{The Quarterly Journal of Economics} \textbf{136}, 1255
(2021).

\bibitem{carvalho_production_2019} V.~M.~Carvalho and A.~Tahbaz-Salehi, \emph{Annual Review of
Economics} \textbf{11}, 635 (2019).

\bibitem{cohen_economic_2008}  L.~Cohen and A.~Frazzini, \emph{The Journal of Finance} \textbf{63}
(2008).

\bibitem{mungo_reconstructing_2023} L.~Mungo, F.~Lafond, P.~Astudillo-Estévez and J.~D.~Farmer, \emph{Journal of Economic Dynamics and Control} \textbf{148}, 104607 (2023).

\bibitem{dhyne_belgian_2015} E.~Dhyne, G.~Magerman, and S.~Rub\`inova, The Belgian production network 2002-2012 , Working Paper 288 (NBB
Working Paper, 2015).

\bibitem{dhyne_trade_2021} E.~Dhyne, A.~K.~Kikkawa, M.~Mogstad, and F.~Tintelnot,
\emph{The Review of Economic Studies} \textbf{88}, 643 (2021).

\bibitem{diem_quantifying_2022} C.~Diem, A.~Borsos, T.~Reisch, J.~Kertész, and
S.~Thurner, \emph{Scientific Reports} \textbf{12}, 7719 (2022).

\bibitem{cardoza_worker_2020}M.~Cardoza, F.~Grigoli, N.~Pierri, and C.~Ruane, IMF
Working paper, No. 20/205 (2020).

\bibitem{chacha_mapping_2022} P.~W.~Chacha, B.~Kirui, and V.~Wiedemann, Mapping Kenya’s Production Network Transaction by
Transaction. Oxford WP

\bibitem{demir_financial_2022} B.~Demir, B.~Javorcik, T.~K.~ Michalski, and E.~Ors, \emph{The Review of Economics and Statistics}, pages 1–46, (2022).

\bibitem{peydro_production_2020} J.~L.~Peydr´o, G.~Jim´enez, H.~Kenan, E.~Moral-Benito, and F.~Vega-Redondo, CEPR Discussion Paper (2020)..

\bibitem{spray_industries_2018} R.~Newfarmer, J.~Page, and F.~Tarp, \textit{Industries without Smokestacks: Industrialization in Africa Reconsidered} (Oxford, 2018)

\bibitem{kumar_distress_2021} A.~Kumar, A.~S.~ Chakrabarti, A.~Chakraborti, T.~Nandi, \emph{Physica A: Statistical Mechanics and its Applications} \textbf{568}, 125714 (2021).

\bibitem{goto_estimating_2017} H.~Goto, H.~Takayasu, and M.~Takayasu, \emph{PLoS ONE} \textbf{12},
e0185712 (2017).

\bibitem{hooijmaaijers_methodology_nodate} S.~Hooijmaaijers and G.~Buiten, OECD Conference, New
Analytical Tools and Techniques for Economic Policy-making (2019).

\bibitem{ialongo_reconstructing_2022} L.~N.~Ialongo, C.~de~Valk, E.~Marchese, F.~Jansen,
H.~Zmarrou, T.~Squartini, and D.~Garlaschelli, \emph{Scientific
Reports} \textbf{12}, 11847 (2022).

\bibitem{inoue_firm-level_2019} H.~Inoue and Y.~Todo, \emph{Nature Sustainability} \textbf{2}, 841
(2019).

\bibitem{inoue_propagation_2020} H.~Inoue and Y.~Todo, \emph{PLoS ONE} \textbf{15}, e0239251 (2020).

\bibitem{kashiwagi_propagation_2021} Y.~Kashiwagi, Y.~Todo, and P.~Matous, \emph{Review of International Economics} \textbf{29}, 1186 (2021).

\bibitem{konig_aggregate_2022} M.~D.~König, A.~Levchenko, T.~Rogers, and F.~Zilibotti,
\emph{Proceedings of the National Academy of Sciences} \textbf{119},
e2203730119 (2022).

\bibitem{kosasih_machine_2022} E.~E.~Kosasih and A.~Brintrup, \emph{International Journal of Production Research} \textbf{60}, 5380 (2022).

\bibitem{maluck_motif_2017} J.~Maluck, R.~V.~Donner, H.~Takayasu, and M.~Takayasu, \emph{Journal of Statistical Mechanics: Theory and Experiment}
 \textbf{2017}, 053404 (2017).

\bibitem{mattsson_functional_2021} C.~E.~S.~Mattsson, F.~W.~Takes, E.~M.~Heemskerk,
C.~Diks, G.~Buiten, A.~Faber, and P.~M.~A.~Sloot, \emph{Frontiers in Big Data} \textbf{4} (2021).

\bibitem{McNerney_how_2021} J.~McNerney, C.~Savoie, F.~Caravelli, V.~M.~Carvalho,
and J.~D.~Farmer, \emph{Proceedings of the National Academy of Sciences} \textbf{119}, e2106031118 (2022).

\bibitem{mizuno_structure_2014} T.~Mizuno, W.~Souma, and T.~Watanabe, \emph{PLoS ONE} \textbf{9}, e100712 (2014).

\bibitem{ohnishi_network_2010} T.~Ohnishi, H.~Takayasu, and M.~Takayasu, \emph{Journal of
Economic Interaction and Coordination} \textbf{5}, 171 (2010).

\bibitem{rachkov_potential_2021} A.~Rachkov, F.~Pijpers, and D.~Garlaschelli, CBS Technical Reports 10.13140/RG.2.2.31861.29925 (2021).

\bibitem{taschereau-dumouchel_cascades_2017} M.~Taschereau-Dumouchel, 2017 Meeting Papers, 700,
Society for Economic Dynamics (2017).

\bibitem{WATANABE2013741} H.~Watanabe, H.~Takayasu, and M.~Takayasu, \emph{Physica
A: Statistical Mechanics and its Applications} \textbf{392}, 741
(2013).


\bibitem{diem_estimating_2023} C.~Diem, A.~Borsos, T.~Reisch, J.~ Kertész and S.~Thurner, \emph{arXiv:2302.11451} (2023).

\bibitem{maluck_network_2015} J.~Maluck and R.~V.~Donner, \emph{PLoS ONE} \textbf{10}, e0133310
(2015).

\bibitem{wang_motif_2022} Z.~Wang, S.~Liu, C.~Han, S.~Huang, X.~Gao, R.~Tang, and Z.~Di, \emph{Frontiers in Physics} \textbf{9} (2022).

\bibitem{alfaro2018costa} A.~Alfaro-Ureña, M.~Fuentes, I.~Manelici, and J.~Vasquez,
Research Paper Series, Banco Central De Costa Rica
(2018).

\bibitem{kito_how_2015}
T.~Kito, S.~New, K.~Ueda,
CIRP Annals 64(1), (2015).

\bibitem{milo_network_2002} R.~Milo, S.~Shen-Orr, S.~Itzkovitz, N.~Kashtan,
D.~Chklovskii, and U.~Alon, \emph{Science} \textbf{298}, 824 (2002).

\bibitem{shen-orr_network_2002} S.~S.~Shen-Orr, R.~Milo, S.~Mangan, and U.~Alon, \emph{Nature
Genetics} \textbf{31}, 64 (2002).

\bibitem{stivala_testing_2021} A.~Stivala and A.~Lomi, \emph{Applied Network Science} \textbf{6}, 1 (2021).

\bibitem{asikainen_cumulative_2020} A.~Asikainen, G.~Iñiguez, J.~Ureña-Carrión, K.~Kaski,
and M.~Kivelä, \emph{Science Advances} \textbf{6}, eaax7310 (2020).

\bibitem{hutchison_triadic_2012} T.~Squartini and D.~Garlaschelli, Self-Organizing Systems \textbf{7166}, 24 (2012).

\bibitem{maratea_triadic_2016} A.~Maratea, A.~Petrosino, and M.~Manzo, \emph{Procedia Computer Science} \textbf{98}, 479 (2016).

\bibitem{squartini_early-warning_2013} T.~Squartini, I.~van Lelyveld, and D.~Garlaschelli, \emph{Scientific Reports} \textbf{3}, 3357 (2013).

\bibitem{squartini_stationarity_2014} T.~Squartini and D.~Garlaschelli, \emph{Journal of Complex Networks} \textbf{3}, 1 (2015).

\bibitem{colomer-de-simon_deciphering_2013} P.~Colomer-de~Simón, M.~Serrano, M.~G.~Beiró, J.~I.
Alvarez-Hamelin, and M.~Boguñá, \emph{Scientific Reports} \textbf{3},
2517 (2013).

\bibitem{jamakovic_how_2009} A.~Jamakovic, P.~Mahadevan, A.~Vahdat, M.~Boguñá,
and D.~Krioukov, \emph{arXiv.0908.1143} (2009).

\bibitem{picciolo_weighted_2022} F.~Picciolo, F.~Ruzzenenti, P.~Holme, and R.~Mastrandrea, \emph{New Journal of Physics} \textbf{24}, 053056 (2022).

\bibitem{jaynes_information_1} E.~T.~Jaynes, \emph{Physical Review} \textbf{106}, 620 (1957).

\bibitem{jaynes_information_1957} E.~T.~Jaynes, \emph{Physical Review} \textbf{108}, 171 (1957).

\bibitem{jaynes_rationale_1982} E.~T.~Jaynes, \emph{Proceedings of the IEEE} \textbf{70}, 939 (1982).

\bibitem{garlaschelli_maximum_2008} D.~Garlaschelli and M.~I.~Loffredo, \emph{Physical Review E} \textbf{78},
015101 (2008).

\bibitem{bardoscia_physics_2021} M.~Bardoscia, P.~Barucca, S.~Battiston, F.~Caccioli,
G.~Cimini, D.~Garlaschelli, F.~Saracco, T.~Squartini, and
G.~Caldarelli, \emph{Nature Reviews Physics} \textbf{3}, 490–507 (2021).

\bibitem{cimini_statistical_2019} G.~Cimini, T.~Squartini, F.~Saracco, D.~Garlaschelli,
A.~Gabrielli, and G.~Caldarelli, \emph{Nature Reviews Physics}
\textbf{1}, 58 (2019).

\bibitem{cimini_mastrandrea_squartini_2021} G.~Cimini, R.~Mastrandrea, and T.~Squartini, Reconstructing Networks, Elements in the Structure and Dynamics of Complex Networks (Cambridge University
Press, 2021).

\bibitem{squartini_garlaschelli_2017} T.~Squartini and D.~Garlaschelli, Maximum-Entropy Networks: Pattern Detection, Network Reconstruction and Graph Combinatorics, SpringerBriefs in Complexity
(Springer Cham, 2017).


\bibitem{garlaschelli_fitness_2004} D.~Garlaschelli and M.~I.~Loffredo, \emph{Physical Review Letters} \textbf{93}, 188701 (2004).

\bibitem{squartini_randomizing_2011} T.~Squartini, G.~Fagiolo, and D.~Garlaschelli, \emph{Physical Review E} \textbf{84}, 046117 (2011).

\bibitem{squartini_randomizing_2011-1} T.~Squartini, G.~Fagiolo, and D.~Garlaschelli, \emph{Physical Review E} \textbf{84}, 046118 (2011).

\bibitem{mastrandrea_enhanced_2014} R.~Mastrandrea, T.~Squartini, G.~Fagiolo, and D.~Garlaschelli, \emph{New Journal of Physics} \textbf{16}, 043022 (2014).

\bibitem{parisi_faster_2020} F.~Parisi, T.~Squartini, and D.~Garlaschelli, \emph{New Journal of Physics} \textbf{22}, 053053 (2020).

\bibitem{Almog_2019} A.~Almog, R.~Bird, and D.~Garlaschelli, \emph{Frontiers in
Physics} \textbf{7}, 10.3389/fphy.2019.00055 (2019).

\bibitem{di_vece_gravity_2022} M.~Di Vece, D.~Garlaschelli, and T.~Squartini, \emph{Physical
Review Research} \textbf{4}, 033105 (2022).

\bibitem{di_vece_reconciling_2023} M.~Di Vece, D.~Garlaschelli, and T.~Squartini, \emph{Chaos,
Solitons \& Fractals} \textbf{166}, 112958 (2023).

\bibitem{cimini_systemic_2015} G.~Cimini, T.~Squartini, D.~Garlaschelli, and A.~Gabrielli,
\emph{Scientific Reports} \textbf{5}, 15758 (2015).

\bibitem{anand_missing_2018} K.~Anand, I.~van Lelyveld, Banai, S.~Friedrich, R.~Garratt, G.~Hałaj, J.~Fique, I.~Hansen, S.~M.~Jaramillo,
H.~Lee, J.~L.~Molina-Borboa, S.~Nobili, S.~Rajan,
D.~Salakhova, T.~C.~Silva, L.~Silvestri, and S.~R.~S.
d.~Souza, \emph{Journal of Financial Stability} \textbf{35}, 107 (2018).

\bibitem{lebacher_in_2019} M.~Lebacher, S.~Cook, N.~Klein, and G.~Kauermann,
\emph{Journal of Network Theory in Finance} \textbf{5}, 29 (2019).

\bibitem{RAMADIAH2020103817} A.~Ramadiah, F.~Caccioli, and D.~Fricke, \emph{Journal of Economic Dynamics and Control} \textbf{111}, 103817 (2020).

\bibitem{mazzarisi_methods_2017} P.~Mazzarisi and F.~Lillo, in Econophysics and Sociophysics: Recent Progress and Future Directions, New Economic Windows, (Springer International Publishing,
Cham, 2017) pp.~201–215.

\bibitem{squartini_analytical_2011} T.~Squartini and D.~Garlaschelli, \emph{New Journal of Physics}
\textbf{13}, 083001 (2011).

\bibitem{squartini_reciprocity_2013} T.~Squartini, F.~Picciolo, and F.~Ruzzenenti, \emph{Scientific
Reports} \textbf{3}, 2729 (2013).

\bibitem{shapiro_analysis_1965} S.~S.~Shapiro and M.~B.~Wilk, \emph{Biometrika} \textbf{52}, 591 (1965).

\bibitem{di_vece_deterministic_2023} M.~Di Vece, D.~Garlaschelli, and T.~Squartini, \emph{Physical
Review E} \textbf{108}, 054301 (2023).

\bibitem{borsos_unfolding_2020} A.~Borsos, M.~Stancsics, \emph{MNB Occasional Papers}, No. 139 (2020).

\end{thebibliography}
\end{document}